\documentclass[lettersize,journal]{IEEEtran}
\usepackage{amsmath,amsfonts}
\usepackage{algorithmic}
\usepackage{array}
\usepackage[caption=false,font=normalsize,labelfont=sf,textfont=sf]{subfig}
\usepackage{textcomp}
\usepackage{stfloats}
\usepackage{url}
\usepackage{verbatim}
\usepackage{graphicx}
\usepackage{float}
\usepackage{booktabs}
\usepackage{cite}
\def\BibTeX{{\rm B\kern-.05em{\sc i\kern-.025em b}\kern-.08em
    T\kern-.1667em\lower.7ex\hbox{E}\kern-.125emX}}
\usepackage{balance}

\begin{document}
\title{Inference skipping for more efficient real-time speech enhancement with parallel RNNs}
\author{Xiaohuai Le, Tong Lei, Kai Chen and Jing Lu
\thanks{Manuscript received ---, 2022; revised ---, 2022; accepted ---, 2022. Date of publication ---, 2022; date of current version ---, 2022.  This work was supported by the National Natural Science Foundation of China under Grant 11874219. The associate editor coordinating the review of this manuscript and approving it for publication was ---. (Corresponding author: ---.)

Xiaohuai Le, Tong Lei, Kai Chen and Jing Lu are with the Key Laboratory of Modern Acoustics, Institute of Acoustics, Nanjing University, Nanjing 210093, China, with the NJU-Horizon Intelligent Audio Lab, Horizon Robotics, Beijing 100094, China, and also with the Nanjing Institute of Advanced Artificial Intelligence, Nanjing 210014, China (e-mail: xiaohuaile@smail.nju.edu.cn; tonglei@smail.nju.edu.cn; chenkai@nju.edu.cn; lujing@nju.edu.cn).

Digital Object Identifier 10.1109/TASLP.2019.2915167}}

\markboth{Journal of \LaTeX\ Class Files,~Vol.~18, No.~9, September~2020}%
{How to Use the IEEEtran \LaTeX \ Templates}

\maketitle

\begin{abstract}
Deep neural network (DNN) based speech enhancement models have attracted extensive attention due to their promising performance. However, it is difficult to deploy a powerful DNN in real-time applications because of its high computational cost. Typical compression methods such as pruning and quantization do not make good use of the data characteristics. In this paper, we introduce the Skip-RNN strategy into speech enhancement models with parallel RNNs. The states of the RNNs update intermittently without interrupting the update of the output mask, which leads to significant reduction of computational load without evident audio artifacts. To better leverage the difference between the voice and the noise, we further regularize the skipping strategy with voice activity detection (VAD) guidance, saving more computational load. Experiments on a high-performance speech enhancement model, dual-path convolutional recurrent network (DPCRN), show the superiority of our strategy over strategies like network pruning or directly training a smaller model. We also validate the generalization of the proposed strategy on two other competitive speech enhancement models.
\end{abstract}


\section{Introduction}
\IEEEPARstart{S}{peech} enhancement (SE) aims at separating clean speech from background interferences for higher speech intelligibility and perceptual quality. Conventional rule-based signal processing algorithms such as OMLSA \cite{ref1} and MMSE-STSA \cite{ref2} are widely used in real-time applications due to their small computational cost, but their performance deteriorates significantly in environments with non-stationary noise. In the last decade, data-driven SE algorithms based on deep neural network (DNN) have achieved remarkable progress with more noise suppression and higher speech quality than the conventional methods. Typical neural networks such as convolutional neural networks (CNNs) \cite{ref3}, recurrent neural networks (RNNs) \cite{ref4} and more recently the attention mechanisms \cite{ref5} have been successfully introduced to time-frequency domain \cite{ref6,ref7} and time domain \cite{ref8,ref9} SE. Generally, RNNs are suitable for real-time processing and attention is more powerful in obtaining contextual information. Compared to other networks, CNNs require fewer parameters through weight sharing mechanism. With a convolutional encoder-decoder (CED) structure and the recurrent bottleneck, the recently proposed convolution recurrent network (CRN) \cite{ref10} can take advantage of both CNNs and RNNs and has become popular in real-time SE \cite{ref11,ref12}.

Most powerful DNNs that achieve promising SE performance require billions of floating-point operations per second. For low latency applications with limited computational resources, we need lightweight DNNs. In addition to designing a small neural network directly \cite{ref13}, compressing a large neural network by quantization \cite{ref14} and pruning \cite{ref15} is also feasible and has been used in SE recently \cite{ref16}. Quantization reduces the bit width of weights and operators for faster inference while pruning removes the less important weights for less resource usage. Knowledge distillation method \cite{ref17} trains a small network under the supervision of a larger network, and this has also been used in SE \cite{ref18}. All these methods are applied at training stage and cannot be utilized to dynamically decrease the computational load according to the characteristics of input data during inference.

Skip-RNN \cite{ref19} dynamically skips the state updates of RNN, which can theoretically reduce the computational load in inference. Some attempts of applying Skip-RNN have been made in SE \cite{ref20}. One notable drawback of this strategy is that the output mask stops updating when the RNN skips, which results in audio artifacts. The exponential smoothing of hidden states \cite{ref20} cannot effectively alleviate the problem. To fully solve this problem, it is necessary to skip the updates without interrupting the updates of mask by leveraging more reasonable network structures. On the other hand, the noisy speech features are often distributed unevenly across time and frequency domain. The noisy signal in practical applications may have the continuous pure noise frames, and they can be exploited to further reduce the computational cost without significant performance degradation.  

We find that the interruption of the mask updates can be effectively circumvented in a network with parallel RNNs when applying the Skip-RNN strategy. Parallel RNNs have been used in various high performance SE models such as the convolutional U-net for speech enhancement (CRUSE) \cite{ref12,ref21}, the gated convolutional recurrent network (GCRN) \cite{ref11} and the dual-path convolutional recurrent network (DPCRN) \cite{ref22}. In this paper, we take DPCRN as an example to discuss the application of Skip-RNN, which can be easily extended to CRUSE and GCRN. DPCRN combines the dual-path RNN (DPRNN) \cite{ref23}, a time domain speech separation network, and CRN in an effective way. Ranked at the 3rd place in Deep Noise Suppression-3 (DNS-3) challenge \cite{ref27}, it has much fewer parameters and lower computational burden than the other top models (0.8 M parameters and 3.7 G MACs, compared with 6.4 M parameters and 6.0 G MACs of Rank-1 model and 5.2 M parameters and 52.5 G MACs\footnote{The computional complexity was not provided in the Rank-2 paper, so we implemented the model and calculated the MACs.} of Rank-2 model), which makes it a very competitive real-time SE method. A huge number of parallel RNNs used in DPCRN contribute significantly to the whole computational load, so it is meaningful to investigate efficient inference skipping strategies. In this paper, we apply Skip-RNN into DPRNN to reduce the computational load during inference. A binary gate is introduced in Skip-RNN to intermittently update the hidden states of RNNs. Due to the utilization of parallel RNNs, the mask will keep updating even when the hidden states stop updating. We investigate several regularization strategies to induce more skipped states while keeping a high quality enhanced speech. The CED structure used in DPCRN can also mitigate audio artifacts. Furthermore, we observe a strong correlation between the skipping and the characteristics of noisy speech spectrogram. In pure noise frames, the states skip mechanism works more frequently. Inspired by this, we introduce a skipping control factor into Skip-RNN and integrate a VAD-guided skipping control method into our model, which can effectively save more computational load.

We test the efficacy of our proposed model by comparing its performance with that of the original DPCRN through extensive experiments. To better illustrate the efficacy of the proposed model, we also compare it with another two models with the same computational load, i.e., a smaller model and the model compressed using network pruning technique. The results of multiple objective metrics show that with the same amount of computation the proposed method achieves significantly better performance. Experiments on other real-time SE models such as CRUSE and GCRN also show that our proposed strategy can significantly reduce the computational load while maintaining competitive performance.

\section{Model description}

\subsection{Problem Formulation}
With the short time Fourier transform (STFT), the noisy speech can be expressed in the T-F domain as
\begin{equation}
	X(t,f)=S(t,f)+N(t,f)\label{eq1}
\end{equation}
where \(X(t,f)\), \(S(t,f)\) and \(N(t,f)\) represent the complex spectra of the noisy speech, the clean speech and the noise, respectively, with time index \(t\) and frequency index \(f\). In order to recover clean speech from the mixture in the time-frequency domain, a common way is to estimate a mask and multiply it by the noisy speech \(X(t,f)\) \cite{ref7}. For phase retrieval, the complex ratio mask (CRM) is a widely used training target \cite{ref24} which is denoted as a complex value \(M_c(t,f)\). In \cite{ref22}, we used CRM to recover the phase implicitly and the denoising process can be expressed as the complex product of the mask and the noisy speech as
\begin{equation}
\widetilde{S}\left(t,f\right)=X\left(t,f\right)\odot M_c\left(t,f\right)\label{eq2}
\end{equation} 
where \(\odot\) denotes element-wise multiplication and \(\widetilde{S}\left(t,f\right)\) is the enhanced speech. Compared with estimating magnitude mask only, CRM is more difficult to learn \cite{ref11}. In this paper, we separately estimate masks for magnitude and phase spectrogram. Similar to PHASEN \cite{ref25}, the model has three outputs, i.e. the magnitude mask \(M\left(t,f\right)\) and the real and imaginary parts of the phase mask \(\mathrm{\Phi}\left(t,f\right)\). Finally, the noisy speech can be enhanced by 
\begin{equation}
\widetilde{S}\left(t,f\right)=X\left(t,f\right)\odot M\left(t,f\right)\odot\mathrm{\Phi}\left(t,f\right){\rm{.}}\label{eq3}
\end{equation}
Instead of estimating the masks directly, we apply the signal approximation (SA) \cite{ref26} to optimize the target masks end-to-end. SA minimizes the difference between the enhanced speech and clean speech with the loss function described as \(\mathcal{L}=Loss\left(\widetilde{S}\left(t,f\right),S\left(t,f\right)\right)\), where \(Loss\) is the complex compressed spectrum MSE loss \cite{ref12} in this paper.

\subsection{Model Architecture}
\subsubsection{Dual-Path Convolution Recurrent Network}
\begin{figure}[tbh!]
	\centering
	
	\includegraphics[width=0.5\textwidth]{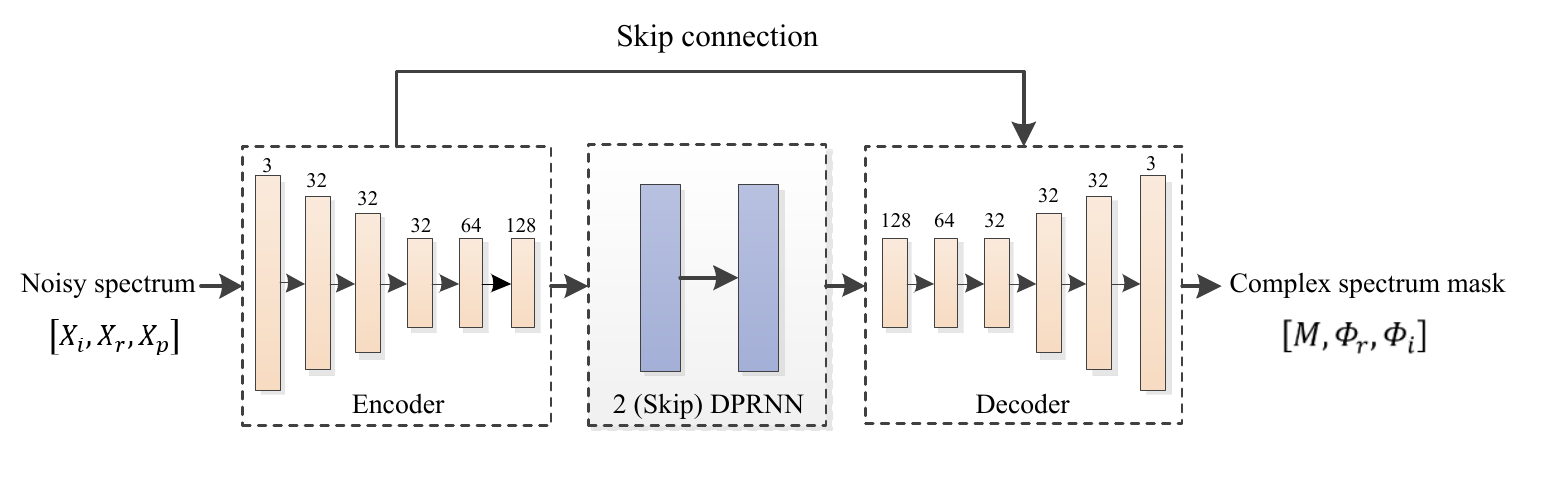}
	
	\caption{The architecture of DPCRN, including the encoder, the DPRNN module and the decoder. The numbers of channels for the convolutional layers are presented.}
	\label{fig:figure1}
	
\end{figure}

\begin{figure}[tbh!]
	\centering
	
	\includegraphics[width=0.5\textwidth]{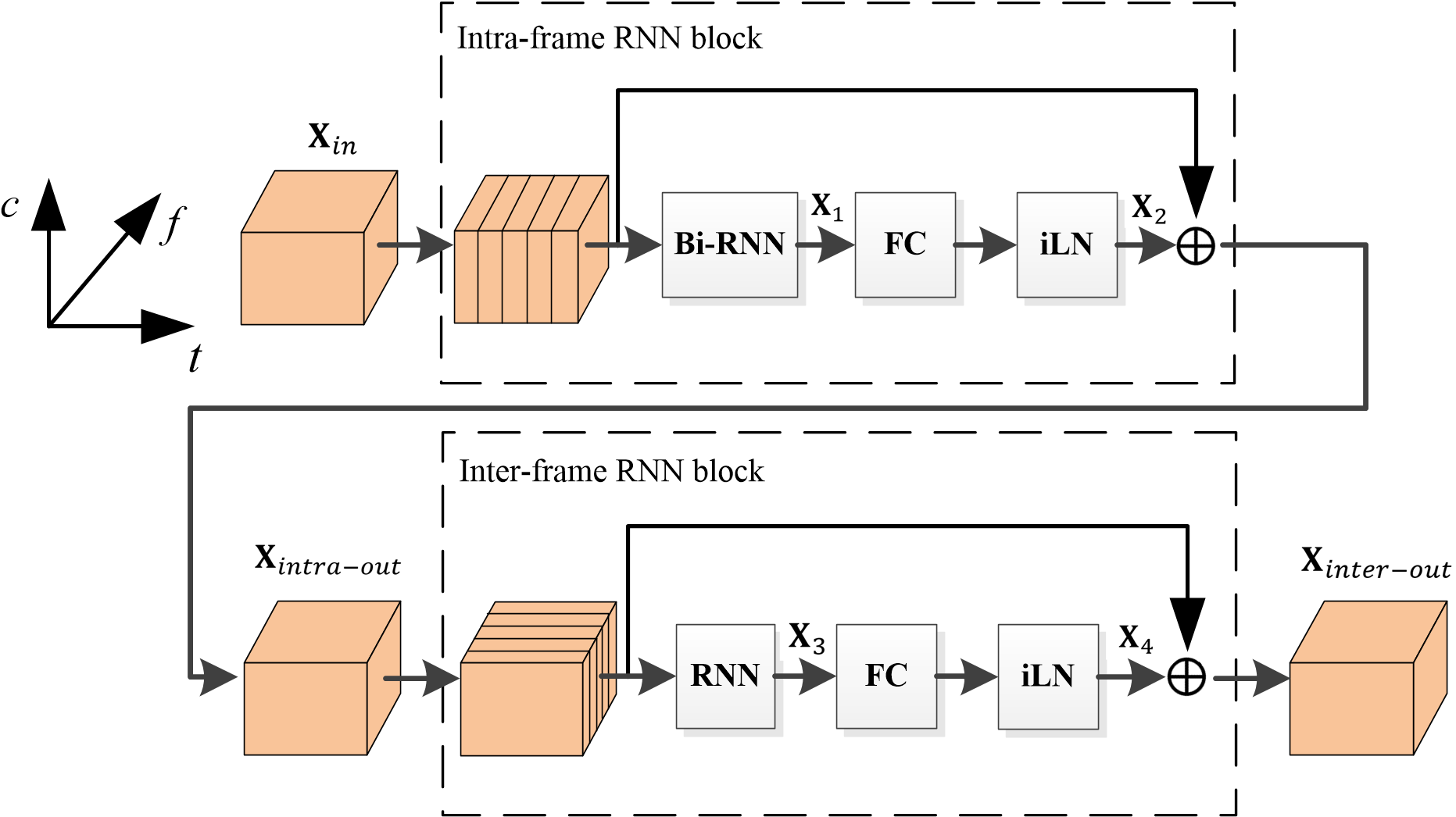}
	
	\caption{The diagram of the DPRNN module. It consists of the intra-frame RNN block and the inter-frame RNN block.}
	\label{fig:figure2}
	
\end{figure}

DPCRN mainly consists of an encoder, a DPRNN \cite{ref23} module and a decoder, as shown in Fig.~\ref{fig:figure1}. The model has three inputs \(X_r\), \(X_i\) and \(X_p\), which respectively denote the real part, the imaginary part of the noisy spectrogram and the logarithmic power spectrogram. They are concatenated as three channels of the input which can be denoted as a 3-D tensor \(\mathbf{X}\in\mathbb{R}^{T\times F\times C}\), where \(T\), \(F\) and \(C=3\) represent the dimensions of frequency, time and channel, respectively. The tensor normalized by batch normalization \cite{ref28} is fed into the encoder to compress the frequency dimension to \(F_{en}\) and expand the channel dimension to \(C_{en}\). The encoder uses the 2-D convolutional (Conv-2D) layers to extract the local patterns from the noisy spectrogram and get the output feature \(\mathbf{X}_{en}\in\mathbb{R}^{T\times F_{en}\times C_{en}}\). In the DPRNN module, the feature is further processed by the intra-frame RNN block and the inter-frame RNN block (also known as the intra-chunk and the inter-chunk RNN block), as depicted in Fig.~\ref{fig:figure2}. The intra-frame RNNs are applied to the \(f\)-dimension of \(\mathbf{X}_{en}\) to model the spectral patterns in a single frame. As for the inter-frame RNNs, parallel RNNs are used to model the time dependence of the \(F_{en}\) sub-features. 

\begin{figure*}[tbh!]
	\centering
	\subfloat[]{
	\begin{minipage}{5.5cm}
		\centering
		\includegraphics[width=1\textwidth]{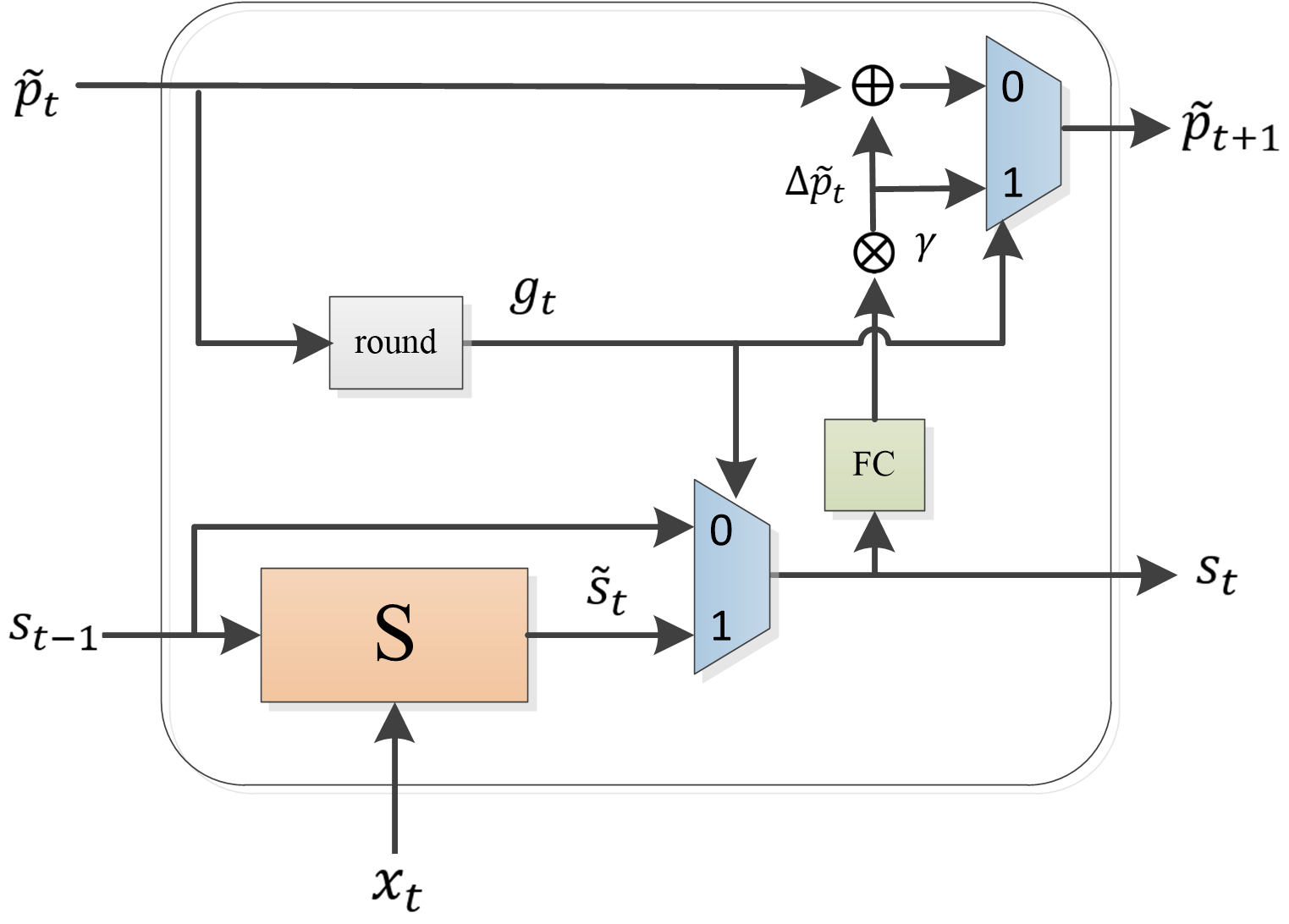}
	\end{minipage}
	}
	\quad
	\subfloat[]{
	\begin{minipage}{5.5cm}
	\centering
	\includegraphics[width=1\textwidth]{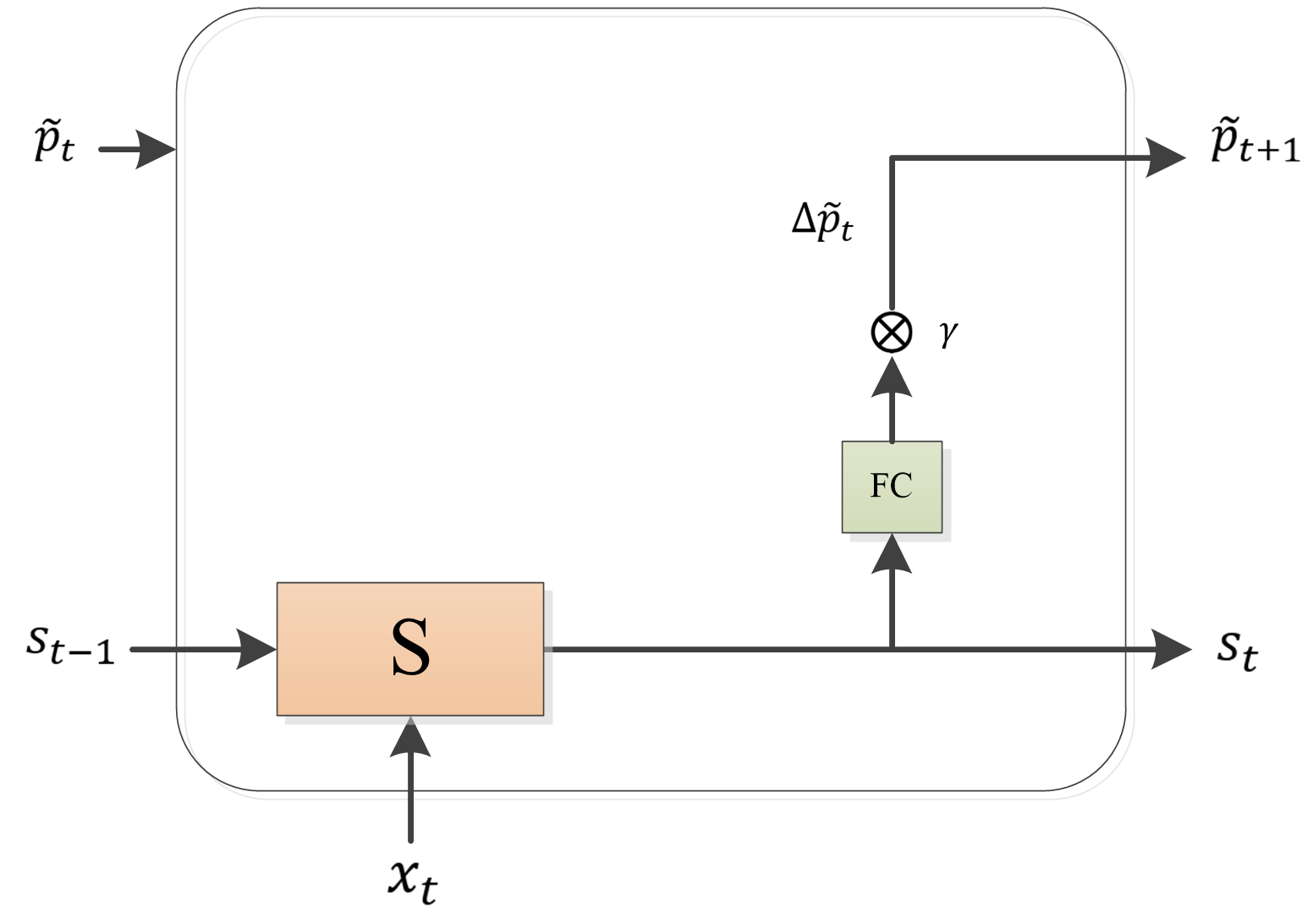}
	\end{minipage}
	}
	\quad
	\subfloat[]{
	\begin{minipage}{5.5cm}
	\centering
	\includegraphics[width=1\textwidth]{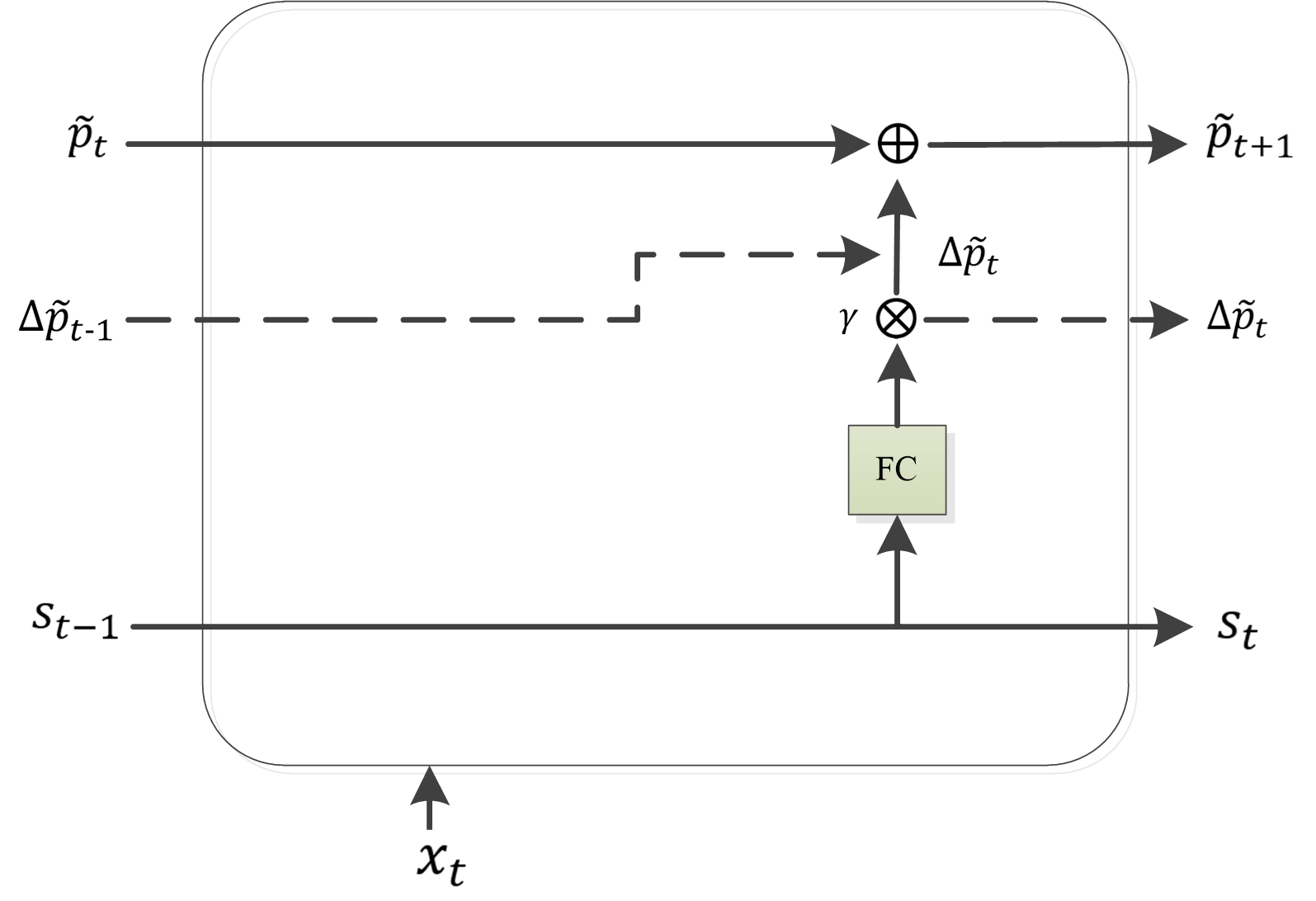}
	\end{minipage}
	}		 
	\caption{The architecture of Skip-RNN with (a) the complete architecture at training, (b) the process of state updating and (c) the process of state skipping.}
	
	\label{fig:figure3}
\end{figure*}

For the input \(\mathbf{X}_{in}{\in\mathbb{R}}^{T\times F_{in}\times C_{in}}\) of a DPRNN block (shown as the orange cube in Fig.~\ref{fig:figure2}), the output of the intra-frame RNN can be expressed as
\begin{equation}
	\mathbf{X}_1=\left[f_{RNN_{intra}}\left(\mathbf{X}_{in}\left[t,:,:\right]\right),t=1,\ldots,T\right]\label{eq4}
\end{equation}
where \(\mathbf{X}_{in}\left[t,:,:\right]\in\mathbb{R}^{F_{in}\times C_{in}}\) is the sub-feature at time index \(t\) and \(f_{RNN_{intra}}\left(\cdot\right)\) represents the function of the intra-frame RNN in the form of a bi-directional RNN with \(C_{hid}/2\) channel output in each direction. The outputs of all frames are concatenated to form a 3-D tensor \(\mathbf{X}_1\in\mathbb{R}^{T\times F_{in}\times C_{hid}}\). A fully connected layer (FC) and the instant layer normalization (iLN) \cite{ref29} are applied on \(\mathbf{X}_1\) to transform the channel dimension back to \(C_{in}\) and normalize the feature respectively, expressed as
\begin{equation}
	\mathbf{X}_2=f_{LN_1}(f_{FC_1}\left(\mathbf{X}_1\right))\label{eq5}
\end{equation}
where \(f_{FC_1}\left(\cdot\right)\) and \(f_{LN_1}\left(\cdot\right)\) respectively denote the functions of FC and iLN. To mitigate the gradient vanishing problem, a residual connection \cite{ref30} is then applied between the input of intra-frame RNN and the output of iLN, expressed as
\begin{equation}
	\mathbf{X}_{intra-out}=\mathbf{X}_2{+\mathbf{X}}_{in}\label{eq6}
\end{equation}
where \(\mathbf{X}_{intra-out}{\in\mathbb{R}}^{T\times F_{in}\times C_{in}}\) is the output of the intra-frame RNN block. Similarly, the inter-frame RNN processes the sub-features at different frequency indices, expressed as 
\begin{equation}
\mathbf{X}_3=\left[f_{RNN_{inter}}\left(\mathbf{X}_{intra-out}\left[:,f,:\right]\right),f=1,\ldots,F_{in}\right]\label{eq7}
\end{equation}
where \(\mathbf{X}_{intra-out}\left[:,f,:\right]\in\mathbb{R}^{T\times C_{in}}\) is the sub-feature at frequency index \(f\) and \(f_{RNN_{inter}}\left(\cdot\right)\) denotes the function of the inter-frame RNN. Similar to the sub-band SE method \cite{ref31}, the sub-features processed by the inter-frame RNN are mapped from certain sub-bands of the original spectrum. The unidirectional RNNs are used to model the time dependence of these sub-features in parallel to guarantee the causality of the whole model. With the FC layer, iLN and the residual connection, the output of the inter-frame RNN block is denoted as \(\mathbf{X}_{inter-out}{\in\mathbb{R}}^{T\times F_{in}\times C_{in}}\), given by
\begin{equation}
\mathbf{X}_4=f_{LN_2}(f_{FC_2}\left(\mathbf{X}_3\right))\label{eq8}	
\end{equation}
\begin{equation}
\mathbf{X}_{inter-out}=\mathbf{X}_4{+\mathbf{X}}_{intra-out}.\label{eq9}
\end{equation}

The decoder uses the transposed convolutional layers to restore the features from the DPRNN module to the original size, forming a symmetric structure with the encoder. There are skip connections between the encoder and the decoder to pass the detailed information and further mitigate the gradient vanishing problem. The output of the decoder also has three channels, the first one is the magnitude mask \(M\) and the remaining two are normalized to obtain the real part and imaginary part of phase mask \(\mathrm{\Phi}\), which are utilized in Eq.~(\ref{eq3}) to enhance the speech. 

In this paper we replace the LSTM in the original DPCRN with GRU \cite{ref32}, which yields little performance loss but saves a quarter of the computational load \cite{ref12}. We also reduce the kernel size of the convolutional layers from (2,5) and (2,3) to (1,5) and (1,3), which reduces the complexity without significant performance loss \cite{ref11}. We use 5 2-D convolutional layers in the encoder and set the strides as \{(2,1), (2,1), (2,1), (1,1), (1,1)\}. Note that the strides in the last two convolutional layers are (1,1) for sufficient frequency resolution of the features fed into the DPRNN module, which we found is important for speech quality \cite{ref22}. We use two DPRNN modules and the resulting baseline DPCRN has 0.53 M trainable parameters. 

The baseline model enhances the noisy speech sampled at 16 kHz with a 32-ms frame length and a 16-ms hop length, resulting in 1115 M multiply-accumulate operations per second (MACs) during inference. Table~\ref{tab:table1} presents the MACs and trainable parameters of each module in the baseline model, from which we can find that the DPRNN module contributes the most to the model complexity. Besides, the convolutional layers can be more easily compressed by the quantization \cite{ref14} technique. Therefore, we focus on the compression of the DPRNN module in this paper.

\begin{table}
	\begin{center}
		\caption{The number of trainable parameters and MACs of each module of the baseline model.}
		\label{tab:table1}
		\begin{tabular}{ c | c | c }
			\toprule
			Operator type & MACs (M) & Parameters (M)\\
			\hline
			Encoder       & 83.71 & 0.042\\
			Intra-frame RNN block \(\times\) 2& 360.6 & 0.180\\ 
			Inter-frame RNN block \(\times\) 2& 458.8 & 0.230 \\
			Decoder & 212.0 & 0.077\\
			\hline 
		    All & 1115.1 & 0.529\\
		    \bottomrule 
		\end{tabular}
	\end{center}
	\vspace{-0.3cm} 
\end{table}

\subsubsection{Skip-RNN}
We apply Skip-RNN \cite{ref19} in the DPRNN module for saving the computational load during inference. The hidden state of the conventional RNN updates in the form as 
\begin{equation}
{\widetilde{s}}_t=S\left(s_{t-1},x_t\right)\label{eq10}
\end{equation}	
where \(S\left(\cdot,\cdot\right)\) is the update function and \(x_t\) is the input at time index \(t\). Skip-RNN introduces a state update probability \({\widetilde{p}}_t\) which is usually mapped to a binary gate \(g_t\in\{0,1\}\) by a round function. The binary gate determines whether the hidden state needs to be updated to \({\widetilde{s}}_t\) or kept the same as the previous step \(s_{t-1}\).
\begin{equation}
	g_t=round\left({\widetilde{p}}_t\right)\label{eq11}
\end{equation}
\begin{equation}
	s_t=g_t{\widetilde{s}}_t+\left(1-g_t\right)s_{t-1}\label{eq12}
\end{equation}
The probability \({\widetilde{p}}_t\) is cumulated by \(\Delta{\widetilde{p}}_t\), which is calculated from the hidden state of the previous step through a FC layer, given by
\begin{equation}
	\Delta{\widetilde{p}}_t=\sigma\left(\mathbf{W}_ps_{t-1}+b_p\right)\label{eq13}
\end{equation}
\begin{equation}
	{\widetilde{p}}_{t+1}=g_t\Delta{\widetilde{p}}_t+(1\ -\ g_t)({\widetilde{p}}_t+min(\Delta{\widetilde{p}}_t,\ \ 1\ -\ {\widetilde{p}}_t))\label{eq14}
\end{equation}
where \(\sigma\) is the Sigmoid function to limit the output between 0 and 1. As shown in Eq.~(\ref{eq14}) and Fig.~\ref{fig:figure3}(a)(b), whenever a state update is skipped, the probability \({\widetilde{p}}_t\) is incremented by \(\Delta{\widetilde{p}}_t\) until \({\widetilde{p}}_t\) is high enough to update the state. As shown by the dashed lines in Fig.~\ref{fig:figure3}(c), \(\Delta{\widetilde{p}}_t\) only depends on the previous state and fixed when the state is fixed, so the update function Eq.~(\ref{eq10}) with the most computational load can be skipped when \(g_t\) is 0. The extra computational load of the additional layer shown in Eq.~(\ref{eq13}) is always negligible, so applying Skip-RNN can significantly reduce the computational load of the whole model. Moreover, for the structure like GRU, since the output is equal to the hidden state, the FC layers following the RNN can also be skipped when the state is fixed. We integrate Skip-RNN in the DPRNN module and get the Skip-intra-RNN and the Skip-inter-RNN respectively as
\begin{equation}
	\mathbf{X}_1^t,\mathbf{g}_1^t=f_{Skip-RNN_{intra}}\left(\mathbf{X}_{in}\left[t,:,:\right]\right)\label{eq15}
\end{equation}
\begin{equation}
	\mathbf{X}_3^f,\mathbf{g}_3^f=f_{Skip-RNN_{inter}}\left(\mathbf{X}_{intra-out}\left[:,f,:\right]\right)\label{eq16}
\end{equation}
\begin{equation}
	\begin{aligned}
		\mathbf{X}_1&=\left[\mathbf{X}_1^t,t=1,\ldots,T\right] \\
		\mathbf{X}_3&=\left[\mathbf{X}_3^f,f=1,\ldots,F_{in}\right]
	\end{aligned}\label{eq17}
\end{equation}
\begin{equation}
	\begin{aligned}
		{\ \mathbf{g}}_1&=\left[\mathbf{g}_1^t,t=1,\ldots,T\right] \\
		{\ \mathbf{g}}_3&=\left[\mathbf{g}_3^f,f=1,\ldots,F_{in}\right]
	\end{aligned}\label{eq18}
\end{equation}
where \(\mathbf{X}_1^t\) and \(\mathbf{X}_3^f\) are the outputs of the Skip-intra-RNN at time index \(t\) and the Skip-inter-RNN at frequency index \(f\), respectively. Consistent with Eqs.~(\ref{eq4}) and~(\ref{eq7}), they are concatenated as the final output \(\mathbf{X}_{1}\) and \(\mathbf{X}_{3}\). \(\mathbf{g}_1^t{\in\mathbb{R}}^{{1\times F}_{in}}\) and \(\mathbf{g}_3^f{\in\mathbb{R}}^{T\times1}\) respectively represent the binary gates of Skip-intra-RNN and Skip-inter-RNN and are also concatenated into \(\mathbf{g}_1{, \mathbf{g}}_3{\in\mathbb{R}}^{{T\times F}_{in}}\) which will be merged into the optimization targets during training. The computational complexity of Skip-RNN can be further measured by the update rate, which is defined as
\begin{equation}
	update{\enspace}rate = \frac{1}{{TF}_{in}}\sum_{t}\sum_{f}{{\ \mathbf{g}}\left[t,f\right]}
	\label{eq19}
\end{equation}
The update rate calculates the percentage of the skipped states over a period of time.

We also propose a method to control the update rate of Skip-RNN to further regularize the computational load during inference. The update of the hidden states of Skip-RNN depends on \({\widetilde{p}}_t\) which is cumulated by \(\Delta{\widetilde{p}}_t\). The increase rate of \({\widetilde{p}}_t\) can be easily adjusted by a scaling factor on \(\Delta{\widetilde{p}}_t\), expressed as
\begin{equation}
	\Delta{\widetilde{p}}_t=\gamma\sigma\left(\mathbf{W}_ps_{t-1}+b_p\right)\label{eq20}
\end{equation}
where \(\gamma\) is the scaling factor, so that the update rate can be controlled. When \(\gamma<1\), the hidden states update rate drops so that more computational load can be saved. This method can be easily implemented during inference without retraining.

\begin{figure}[tbh!]
	\centering
	
	\includegraphics[width=0.5\textwidth]{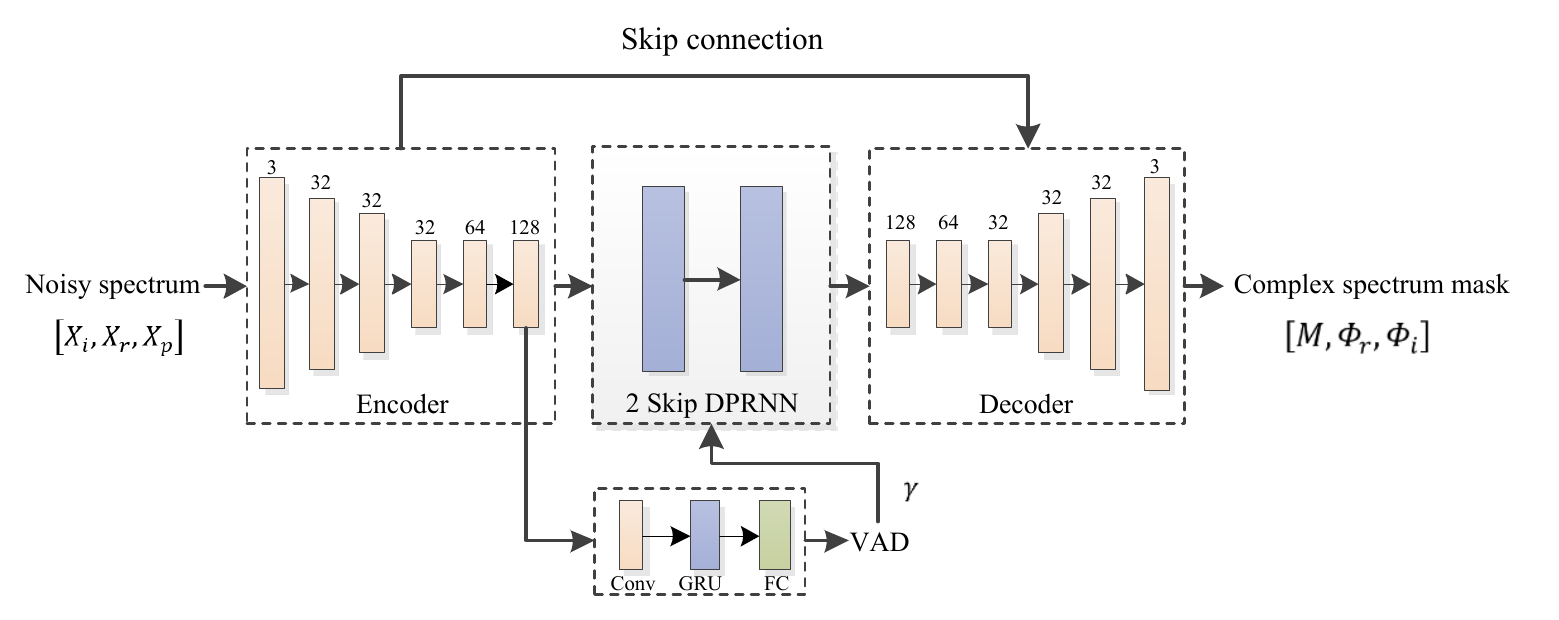}
	
	\caption{The architecture of DPCRN with the Skip-RNN and the VAD estimator.}
	\label{fig:figure4}
	
\end{figure}

To better exploit the characteristic of the signal spectrogram, we also integrate the voice activity detection (VAD) in our model under a multi-task learning framework. As shown in Fig.~\ref{fig:figure4}, the feature output from the encoder is processed by a \(1\times1\) convolutional layer, after which the channel is compressed to 1. A GRU and a FC layer are then applied. The output of the FC layer is the probability of voice formed by the Sigmoid function. The dimension of the hidden states in the GRU is set to 32. The VAD estimator is optimized by minimizing the binary cross-entropy between the true labels and the predictions. Note that the VAD estimator is easy to be trained since we only require a high recall ratio of the voice frame. According to the estimated VAD labels, Eq.~(\ref{eq20}) will be utilized with different \(\gamma\) on the pure noise frames and the voice frames for further computational load adjustment.

\subsection{Training Target}
The complex compressed spectrum MSE loss \cite{ref12} we use can be expressed as
\begin{equation}
	\mathcal{L}_{SD}={\lambda}MSE\left(S^c,{\hat{S}}^c\right)+(1-\lambda)MSE(|{S|}^c,{|\hat{S}|}^c)\label{eq21}
\end{equation}
where \(S=STFT\left(s\right)\) and \(\hat{S}=STFT\left(\hat{s}\right)\) denote the spectrogram of the clean speech \(s\) and the enhanced speech \(\hat{s}\), and \(MSE\left(\cdot\right)\) is the function of the mean squared error (MSE). Note that \(s\) and \(\hat{s}\) in a batch are normalized by the same factor for better optimization. In the first term of \(\mathcal{L}_{SD}\), the spectrogram is compressed by
\begin{equation}
	S^c=|{S|}^cexp(j\theta_S)\label{eq22}
\end{equation}
where \(\theta_S\) is the phase of \(S\). In this paper, the spectrogram compression factor \(c\) and the loss weight factor \(\lambda\) are respectively set to 0.3 and 0.1 for higher speech quality \cite{ref21}.

We observe that directly training the model using Skip-RNN with loss function Eq.~(\ref{eq21}) cannot lead to a high skipping rate, so we use a regularization term to induce more skipped states in each Skip-RNN, expressed as  
\begin{equation}
	\mathcal{L}_{Skip}=\frac{1}{{TF}_{in}}\sum_{i}{ \left(\sum_{t}\sum_{f}{{\ \mathbf{g}}_{1;i}\left[t,f\right]}+\sum_{t}\sum_{f}{{\ \mathbf{g}}_{3;i}\left[t,f\right]}\right)\ }\label{eq23}
\end{equation}
where \(\mathbf{g}_{1;i}\) and \(\mathbf{g}_{3;i}\) represent the binary gates of the Skip-intra-RNN and the Skip-inter-RNN in the \(i\)-th Skip-DPRNN module, respectively. The regularization term minimizes the sum of the mean values of the binary gates for higher compression rate. To obtain a specified compression rate, we can also optimize the states update rate of each Skip-RNN to the target rate by
\begin{equation}
	\begin{aligned}
		\mathcal{L}_{Skip} & =\sum_{i}{\left( \left(\frac{1}{{TF}_{in}}\sum_{t}\sum_{f}{{\ \mathbf{g}}_{1;i}\left[t,f\right]} - \mu_{1;i}\right)^2 + \right.} \\
		&{\left. \left(\frac{1}{{TF}_{in}}\sum_{t}\sum_{f}{{\ \mathbf{g}}_{3;i}\left[t,f\right]} - \mu_{3;i}\right)^2 \right)} 
	\end{aligned}\label{eq24}
\end{equation}
where \(\mu_{1;i}\) and \(\mu_{3;i}\) are the target update rate of the Skip-intra-RNN and the Skip-inter-RNN in the \(i\)-th Skip-DPRNN module, respectively. In addition to MSE, the mean absolute error (MAE) can also be used as a regularization term, given by
\begin{equation}
	\begin{aligned}
		\mathcal{L}_{Skip} & =\sum_{i}{\left( \left|\frac{1}{{TF}_{in}}\sum_{t}\sum_{f}{{\ \mathbf{g}}_{1;i}\left[t,f\right]} - \mu_{1;i}\right| + \right.} \\
		&{\left. \left|\frac{1}{{TF}_{in}}\sum_{t}\sum_{f}{{\ \mathbf{g}}_{3;i}\left[t,f\right]} - \mu_{3;i}\right| \right)} 
	\end{aligned}\label{eq25}
\end{equation}
Compare with MSE, MAE can accelerate the convergence. The final loss function can be expressed as a weighted sum of \(\mathcal{L}_{SD}\) and \(\mathcal{L}_{Skip}\), given as
\begin{equation}
	{\mathcal{L}=\mathcal{L}}_{SD}+\alpha\mathcal{L}_{Skip}\label{eq26}
\end{equation}
where \(\alpha\) is a weighting factor.

\section{Experiments}

\subsection{Datasets}
All the experiments are conducted on the WSJ0 SI-84 dataset \cite{ref33}. We select 14633 utterances (about 25 h) from 120 speakers, including 12776 in training set, 1206 in validation set and 651 in test set. In order to improve the generalization on various noises, we also use the noise dataset from the DNS-3 Challenge \cite{ref27}. The dataset also includes 60000 audio clips from DEMAND \cite{ref34}, Audioset \cite{ref35} and Freesound\footnote{https://freesound.org/} for a total of 150 hours.

To improve the robustness against the variation of the signal levels, the reverberation and the coloring effects of speech signal in practical applications, we use the data augmentation pipeline from \cite{ref36} in training. Specifically, the clean speech is split into 11212 8-second segments, 30\% of which will be filtered by a second-order IIR spectral augmentation filters. Consistent with \cite{ref12,ref13}, all the filter coefficients are uniformly distributed in [-3/8, 3/8] . After spectral augmentation, half of the audio clips are convolved with room impulse responses (RIRs) randomly-selected from openSLR26 and openSLR28 \cite{ref37}. Then the noisy speech is generated by mixing reverberant speech and noise. The signal-to-noise ratio (SNR) of the mixture is randomly sampled between -5 and 5 dB. Finally, the noisy and clean speech is scaled by the same factor drawn from a Gaussian distribution on the denary logarithmic scale with mean -0.5 and variance 1. The level scaling factor will also be used in Eq.~(\ref{eq21}) for normalization. The validation set is generated in the same way as the training set.

For the test set, we use two unseen noise datasets. One is the music data from MUSAN \cite{ref38}, the other includes the babble, factory1 and f16 noise from NOISEX92 \cite{ref39}. The noises are mixed with the speech at SNR of -5, 0 and 5 dB, with 651 mixtures for each SNR level and each dataset. Considering the audio with long pure noise frames in practical applications, we also generate a test set of long noisy speech. We concatenate the test speech in pairs, and there are 8-second silent audio between every pair. We randomly mix the long speech and music noises at the SNR uniformly sampled between -5 to 5 dB. In total, we generate 3906 segments of short audio and 325 segments of long audio for test\footnote{The source code of our proposed model can be found at https://github.com/Le-Xiaohuai-speech/SKIP-DPCRN}.

\subsection{Experiment Configuration}

The window length and hop size used in our models are 32 ms and 16 ms respectively, resulting in a total latency of 48 ms. The FFT length is 512 and the frequency dimension \(F\) of the input is 257. The sine window is applied before FFT and overlap-add. The baseline model is called DPCRN-base, in which the hidden dimension \(C_{hid}\) of every GRU and Bi-GRU is 128. We reduce \(C_{hid}\) to control the computational complexity of DPCRN-base for comparison in two ways. One is by directly training a smaller model, named as DPCRN-S, and the other is by applying the structured pruning method described in \cite{ref40}, named as DPCRN-P, with the intrinsic sparse structures-based \(\ell_2\)-norm integrated in Eq.~(\ref{eq21}) as the loss function. 

We apply Skip-RNN in the intra-frame RNNs and the inter-frame RNNs, and name the resulting models DPCRN-Intra-skip and DPCRN-Inter-skip respectively. In order to explore the compression capacity of Skip-RNN, we also replace all the RNNs with Skip-RNNs and name the resulting model DPCRN-All-skip. The models are trained by minimizing Eq.~(\ref{eq26}) with one regularization term from Eqs.~(\ref{eq23}) and (\ref{eq24}), as explicitly described in Sec. IV.

Based on DPCRN-All-skip, we further explore the multi-task learning framework with a VAD estimator. The ideal VAD labels are generated using energy thresholds of the clean speech. During inference, the output \(\widetilde{v}(t)\) of the estimator is smoothed by the rules expressed as  
\begin{equation}
	{v(t)} = \begin{cases}
		0,&{\text{if}}\ \widetilde{v}(t),\widetilde{v}(t-1),\cdots,\widetilde{v}(t-9)=0 \\ 
		{1,}&{\text{otherwise.}} 
	\end{cases}
	\label{eq27}
\end{equation}
where \(v(t)\) is the final VAD estimation of the \(t\)-th frame. The smoothing strategy improves the recall ratio of the voice frames at the expense of VAD accuracy. 

All the models are trained by the Adam optimizer \cite{ref41} with a batch size of 16. The initial learning rate is 1e-3 and it will be halved if the loss on the validation set does not improve for 10 consecutive epochs. Early stopping is also applied in training if the loss on the validation set does not improve for 20 epochs. For structured pruning, the loss function with the \(\ell_2\)-norm regularization is used first, then the weights of RNNs and corresponding FCs are pruned by their magnitude, after which the models are fine-tuned for 50 epochs with the original loss of Eq.~(\ref{eq21}). TensorFlow is employed for model implementation and a Nvidia TITAN Xp is used for training.

\linespread{1.15}
\begin{table*}[!htbp]
	\caption{Performance of our proposed models on two test sets.}
	\label{tab:table2}
	\centering
	\begin{tabular}{cc c cccc c cccc}
		\toprule
		\multicolumn{2}{c}{Test set} & \textbf{}  & \multicolumn{4}{c}{NOISEX} & \textbf{} & \multicolumn{4}{c}{MUSAN}  \\
		\cline{1-2}  \cline{4-7} \cline{9-12} 
		Models  &  $\alpha$& & SDR (in dB) & PESQ & STOI (in \%) & Update rate & & SDR (in dB) & PESQ & STOI (in \%) & Update rate\\
		\midrule
		Noisy      &   -  && 0.091    & 1.433    & 74.94 & - &  & 0.043 & 2.004 & 80.76 & -   \\
		DPCRN-base &   -  && \textbf{9.527}  & \textbf{2.688} & \textbf{89.13}   & 1.00        & & \textbf{11.54}    & \textbf{3.105} & \textbf{93.69}   & 1.00       \\	
		\hline	
		Intra-skip & $5 \times 10^{-5}$ && 9.508 & 2.681 & 88.87  & 0.712        &                 & 11.53       & 3.084 & 93.55 & 0.679                              \\
		Intra-skip & $1 \times 10^{-4}$ && 9.448 & 2.670 & 88.86  & 0.482        &                 & 11.48       & 3.076 & 93.54 & 0.456                              \\
		Intra-skip & $2 \times 10^{-4}$ && 9.394 & 2.672 & 88.76  & 0.436        &                 & 11.37       & 3.065 & 93.41 & 0.419                              \\
		Intra-skip & $3 \times 10^{-4}$ && 9.358 & 2.649 & 88.49  & \textbf{0.271} &               & 11.24       & 3.042 & 93.17 & \textbf{0.258}                     \\
		\hline
		Inter-skip & $5 \times 10^{-5}$ && 9.409 & 2.666 & 88.67  & 0.765        &                 & 11.45       & 3.068 & 93.38 & 0.709                              \\
		Inter-skip & $1 \times 10^{-4}$ && 9.424 & 2.662 & 88.63  & 0.640        &                 & 11.39       & 3.059 & 93.30 & 0.545                              \\
		Inter-skip & $2 \times 10^{-4}$ && 9.355 & 2.633 & 88.24  & 0.426        &                 & 11.26       & 3.031 & 93.01 & 0.393                              \\
		Inter-skip & $3 \times 10^{-4}$ && 9.355 & 2.640 & 88.30  & 0.359        &                 & 11.27       & 3.024 & 93.02 & 0.328                              \\
		\hline
		All-skip   & $5 \times 10^{-5}$ && 9.405 & 2.654 & 88.57  & 0.745        &                 & 11.42       & 3.056 & 93.29 & 0.706                              \\
		All-skip   & $1 \times 10^{-4}$ && 9.337 & 2.645 & 88.53  & 0.577        &                 & 11.29       & 3.014 & 93.20 & 0.524                              \\
		\bottomrule
	\end{tabular}
\end{table*}

\subsection{Models for Comparison and Evaluation Metrics}
Apart from DPCRN, we also apply the proposed Skip-RNN strategy to two more real-time SE models, i.e., CRUSE \cite{ref12,ref21} and GCRN \cite{ref11}, to further validate its efficacy. CRUSE is a well-studied model which achieves a promising tradeoff between computational load and performance. GCRN uses a gating mechanism and group RNNs, achieving competitive performance in real-time SE. We build the CRUSE model in the same way as \cite{ref21}, which has 8 convolutional layers and a GRU bottleneck with 4 parallel subgroups. As for GCRN, we follow the same structure as \cite{ref11}, but replace LSTMs with GRUs. The number of parameters of these two models are 8.58 M and 7.66 M, respectively. The frame length and hop size in both models are 20 ms and 10 ms. We use the loss function Eq.~(\ref{eq26}) with the regularization Eq.~(\ref{eq25}) to train these models, because we find MAE leads to faster convergence than MSE when the hidden size of RNN is large.

We use three objective metrics for performance evaluation: the signal-to-distortion ratio (SDR) \cite{ref42}, the perceptual evaluation of speech quality (PESQ) \cite{ref43}, and the short-time objective intelligibility (STOI) \cite{ref44}.

\section{Results and discussions}
\subsection{The Efficacy of Skip-RNN}
\linespread{1.15}
\begin{table}
	\begin{center}
		\caption{MACs, trainable parameter numbers of proposed models and the average update rate of each RNN block.}
		\label{tab:table3}
		\setlength{\tabcolsep}{1.5mm}{
		\begin{tabular}{ c | c | c | c | c | c }
			\toprule
			Models & $\alpha$ & {\begin{tabular}[c]{@{}c@{}} Intra-RNN\\update rate\end{tabular} } & {\begin{tabular}[c]{@{}c@{}} Inter-RNN\\update rate\end{tabular} } & {\begin{tabular}[c]{@{}c@{}} MACs\\(M)\end{tabular} }& {\begin{tabular}[c]{@{}c@{}} Para.\\(M)\end{tabular} } \\
			\hline
			DPCRN-base & - & 1.00 & 1.00 & 1115 & \textbf{0.5286} \\
			\hline
			Intra-skip & $5 \times 10^{-5}$ & 0.696 & 1.00 & 1005 & 0.5288\\ 
			Intra-skip & $1 \times 10^{-4}$ & 0.469 & 1.00 & 923.6 & 0.5288 \\
			Intra-skip & $2 \times 10^{-4}$ & 0.427 & 1.00 & 908.6 & 0.5288 \\
			Intra-skip & $3 \times 10^{-4}$ & \textbf{0.265} & 1.00 & 849.7 & 0.5288 \\
			\hline 
			Inter-skip & $5 \times 10^{-5}$ & 1.00 & 0.737 & 1011 & 0.5288 \\
			Inter-skip & $1 \times 10^{-4}$ & 1.00 & 0.575 & 994.4 & 0.5288 \\
			Inter-skip & $2 \times 10^{-4}$ & 1.00 & 0.410 & 844.2 &  0.5288\\
			Inter-skip & $3 \times 10^{-4}$ & 1.00 & \textbf{0.344} & 814.0 & 0.5288 \\
			\hline 
			All-skip   & $5 \times 10^{-5}$ & 0.702 & 0.749 & 892.5 & 0.5291 \\
			All-skip   & $1 \times 10^{-4}$ & 0.520 & 0.581 & \textbf{749.8} & 0.5291 \\
			\bottomrule 
		\end{tabular}
		}
	\end{center}
\end{table}

To demonstrate the efficacy of Skip-RNN, we compare the performance of DPCRN-base and the model with Skip-RNN. Table~(\ref{tab:table2}) presents the performance and the average states update rate on two test sets. All the models with Skip-RNN are trained by minimizing the loss function Eq.~(\ref{eq26}) with the regularization Eq.~(\ref{eq23}) weighted by \(\alpha\). Table~(\ref{tab:table3}) shows the average states update rate, the average MACs and the number of parameters of every model. Note that the average update rate of Skip-RNN slightly differs between different test sets, so we show the average update rate of all the test sets in Table~\ref{tab:table2}.

From Tables~\ref{tab:table2} and~\ref{tab:table3}, one can get the following conclusions. Firstly, integrating Skip-RNN saves significant computational load with limited performance degradation. Secondly, it can also be found that the performance is positively correlated with the amount of computation. Note that although DPCRN-Intra-skip obtains lower update rate than DPCRN-Inter-skip with the same \(\alpha\), its overall computational load shown in Table~\ref{tab:table3} is higher because the amount of computation of an intra-frame RNN is a quarter fewer than an inter-frame RNN. Therefore, DPCRN-Intra-skip gets higher objective metric scores than DPCRN-Inter-skip with the same \(\alpha\). Thirdly, as \(\alpha\) increases, the regularization term leads to more frequent skipping and lower update rate, which can be considered as a form of dynamic temporal pruning. The states update rate relates to both the regularization factor \(\alpha\) and the noise characteristics during inference. Specifically, the performance on the MUSAN test set is superior to the NOISEX test set in terms of all the metrics but the update rate is lower, demonstrating that the model will update more frequently under more disruptive noises and “slack off” under more tractable noises. This makes it possible to further reduce the computational load. Lastly, although DPCRN-All-skip has overall inferior performance, it exceeds two DPCRN-Inter-skip models with fewer MACs when $\alpha = 1 \times 10^{-4}$, which indicates applying Skip-RNN on all RNNs with a proper \(\alpha\) can achieve a good balance between performance and computation. In addition to the results of the average computational load, the distributions of the frame level update rates of the DPRNN module in DPCRN-All-skip ($\alpha = 1 \times 10^{-4}$) are shown in Fig.~\ref{fig:figure5}. The frame-level computational load is distributed like a Gaussian distribution and the peek computational load is reduced by about 15\%.

\begin{figure*}[tbh!]
	\centering
	\subfloat[]{
		\begin{minipage}{6cm}
			\centering
			\includegraphics[width=1\textwidth]{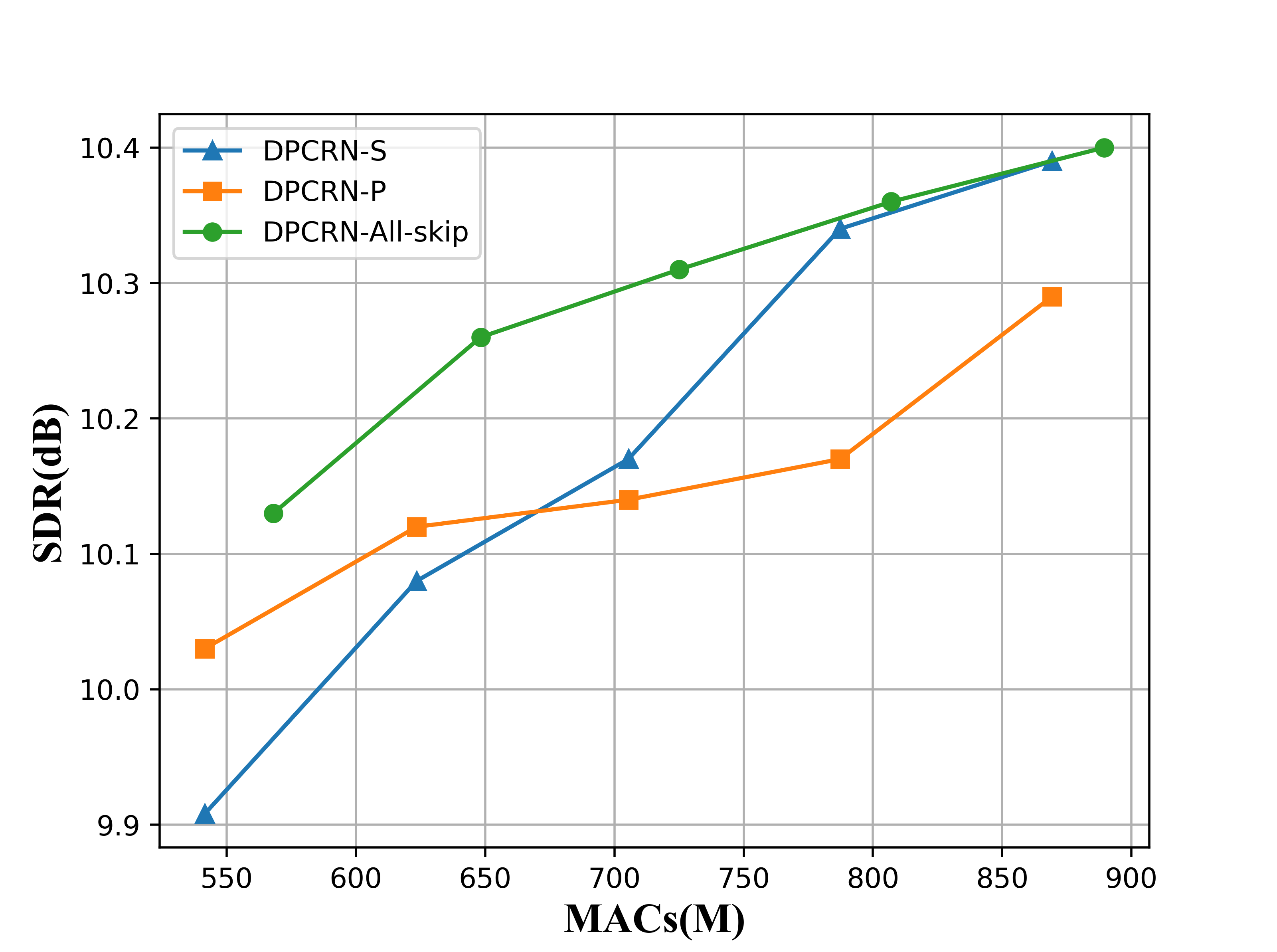}
		\end{minipage}
	}
	\subfloat[]{
		\begin{minipage}{6cm}
			\centering
			\includegraphics[width=1\textwidth]{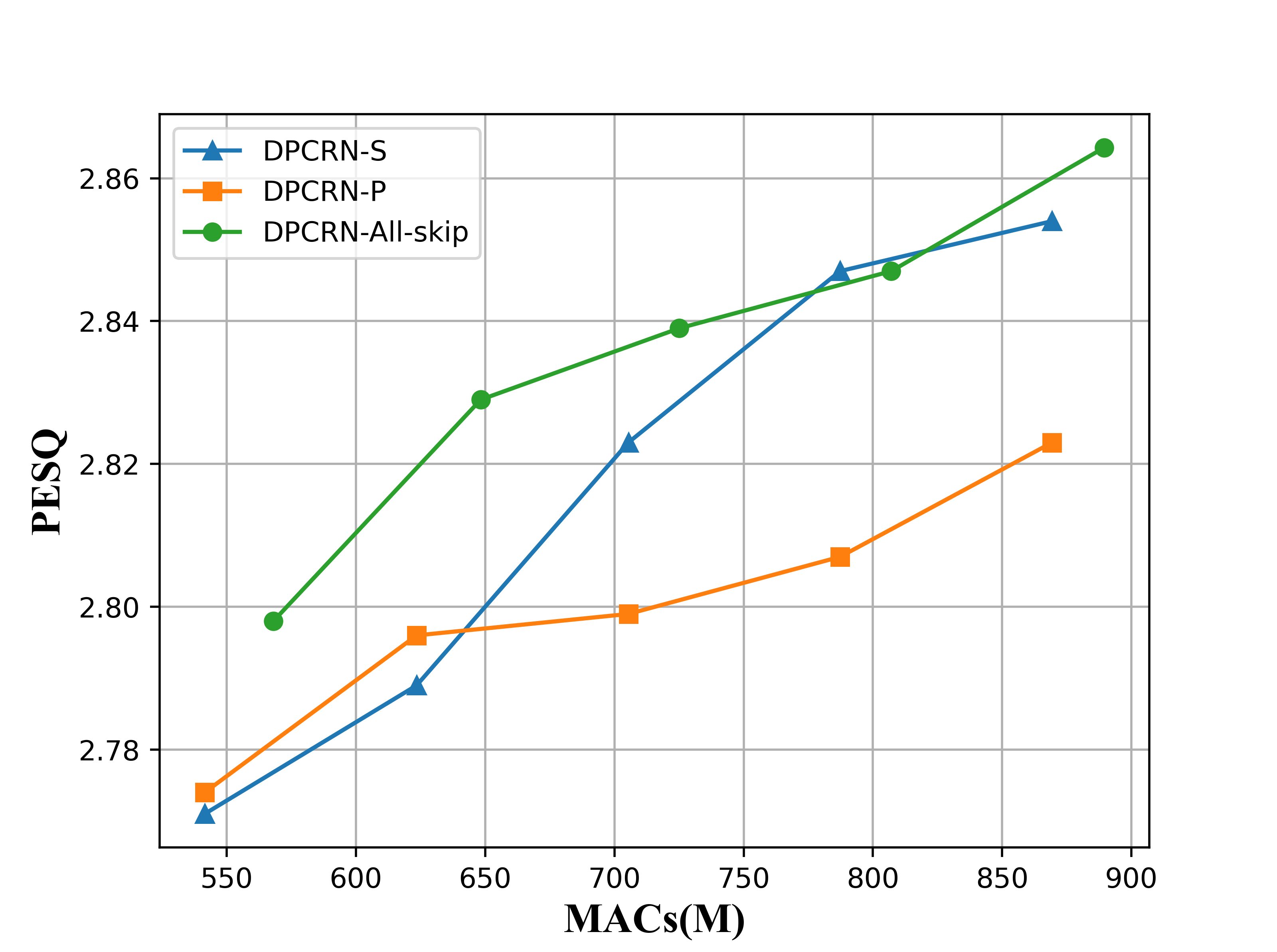}
		\end{minipage}
	}
	\subfloat[]{
		\begin{minipage}{6cm}
			\centering
			\includegraphics[width=1\textwidth]{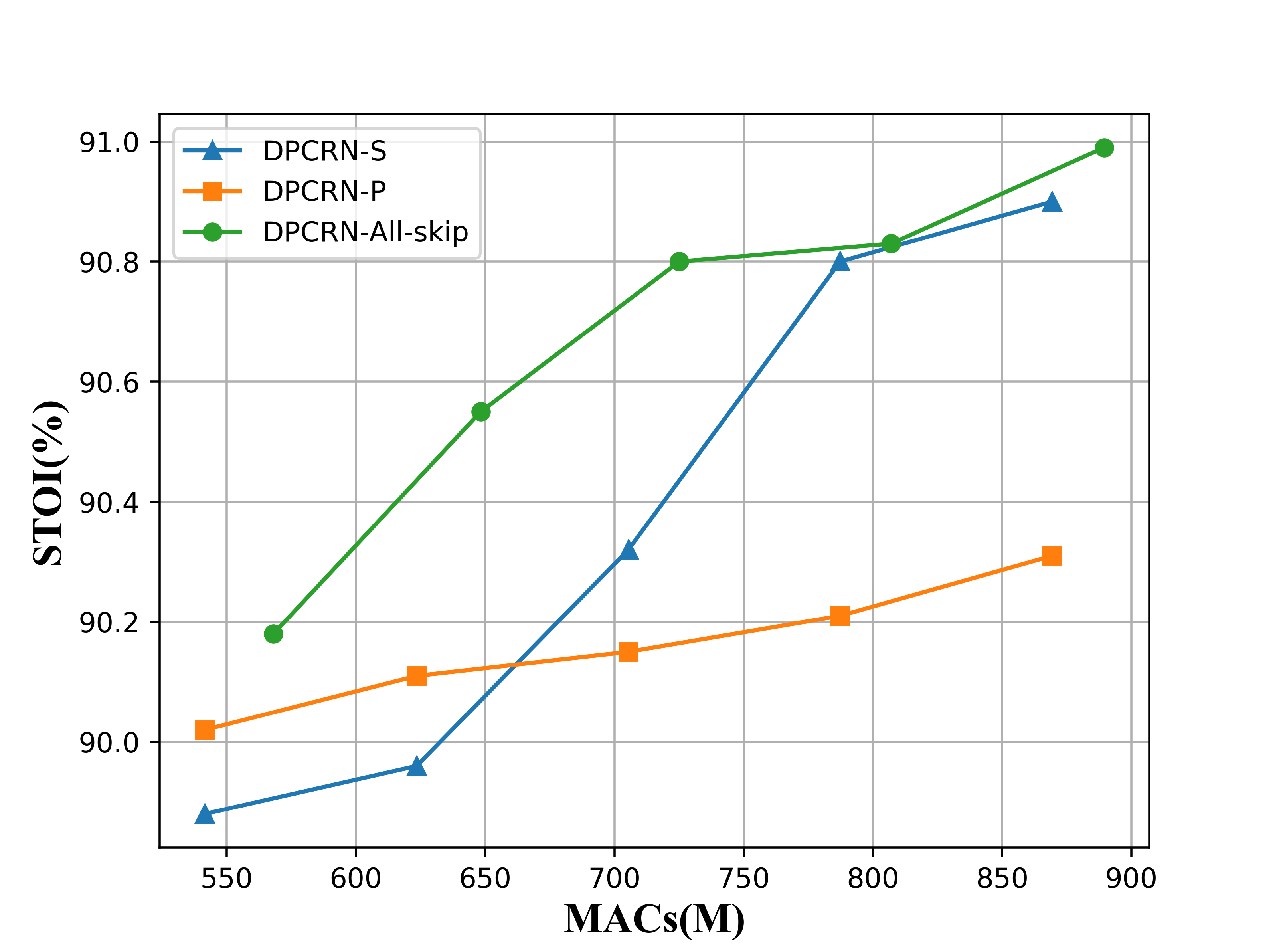}
		\end{minipage}
	}		 
	\caption{The average performance of DPCRN-All-skip with update rates of \{0.3, 0.4, 0.5, 0.6, 0.7\} and DPCRN-S and DPCRN-P with compression rates of \{0.3, 0.4, 0.5, 0.6, 0.7\}.}
	
	\label{fig:figure6}
\end{figure*}

\begin{figure}[tbh!]
	\centering
	
	\includegraphics[width=0.4\textwidth]{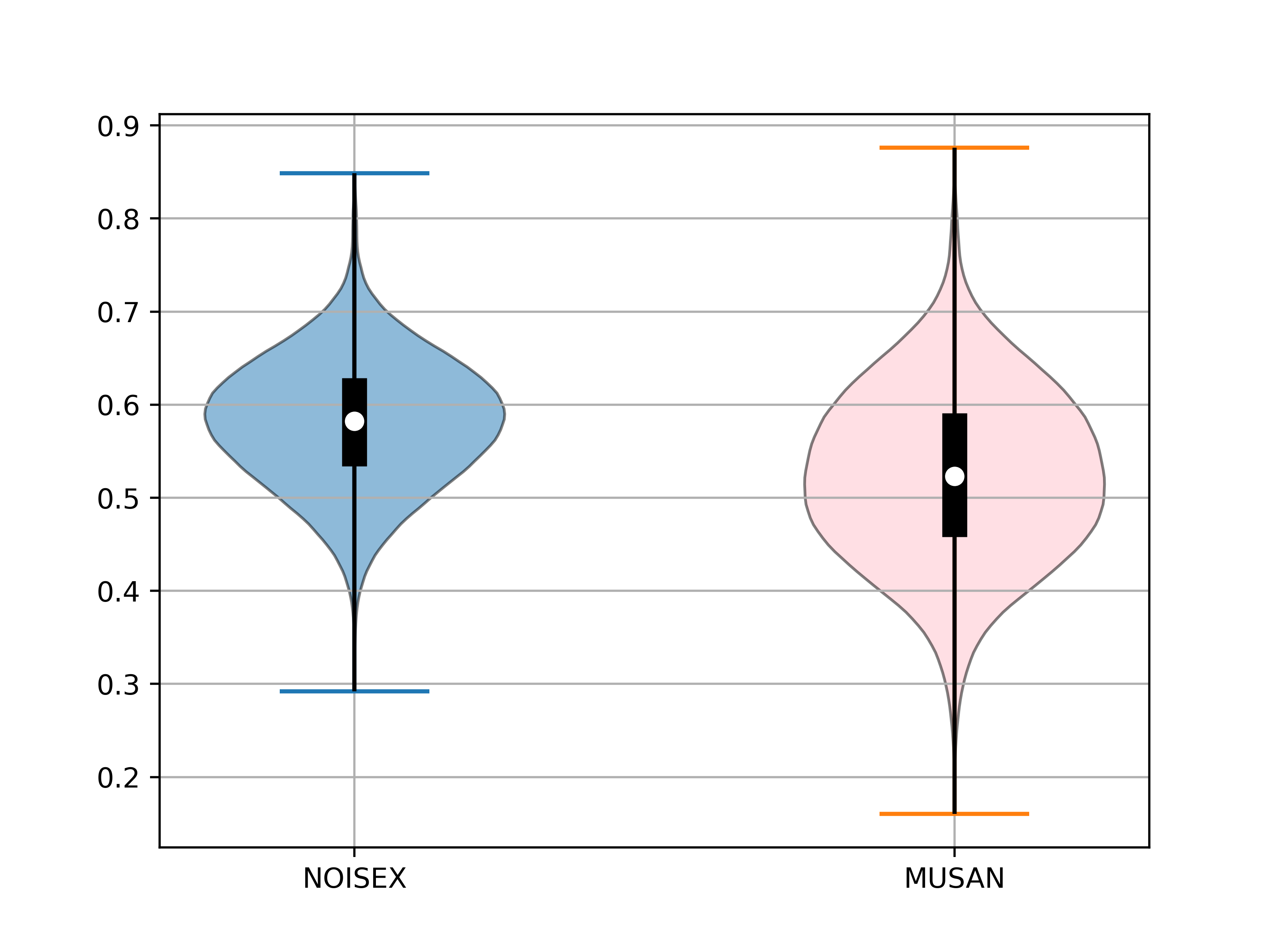}
	
	\caption{The distributions of the frame level update rates of the DPRNN module in DPCRN-All-skip ($\alpha = 1 \times 10^{-4}$).}
	\label{fig:figure5}
	\vspace{-0.5cm}	
\end{figure}

\subsection{Compression Rate Comparison}

To further illustrate the compression performance of the Skip-RNN strategy, Eq.~(\ref{eq26}) with Eq.~(\ref{eq24}) are used as the loss ($\alpha = 1 \times 10^{-2}$) to train the DPCRN-All-skip models with the target update rate respectively set to \{0.3, 0.4, 0.5, 0.6, 0.7\}. For comparison, we obtain DPCRN-S by directly training a smaller model and DPCRN-P using structured pruning. The hidden dimensions of the intra-frame and the inter-frame RNNs of both models are reduced to \{(48,52), (60,65), (74,77), (86,89), (96,100)\}, leading to the compression rates of \{0.3, 0.4, 0.5, 0.6, 0.7\}, so that all the models can be compared with similar computational load. The results of the three methods are shown in Fig.~\ref{fig:figure6}. The average computational load shown by the green line is always higher than expectation, indicating that the achieved update rates of DPCRN-All-skip on the test sets are always a little higher than the target update rates set in training. Even so, it still can be seen that DPCRN-All-skip outperforms the other two models in most cases. When the compression rate is greater than about 0.6, the performance of DPCRN-S is comparable with that of DPCRN-All-skip. DPCRN-P performs worst at a higher compression rate, but with decreasing compression rate, it gradually exceeds DPCRN-S. A possible reason is that the regularization term of the pruning method brings significant weights sparsity, which makes it difficult to fine-tune the model at a higher compression rate.

\subsection{More Efficient Inference by VAD-guided Skipping}

\begin{figure*}[htb!]
	\centering
	\subfloat[]{
		\begin{minipage}{9cm}
			\centering
			\includegraphics[width=1\textwidth]{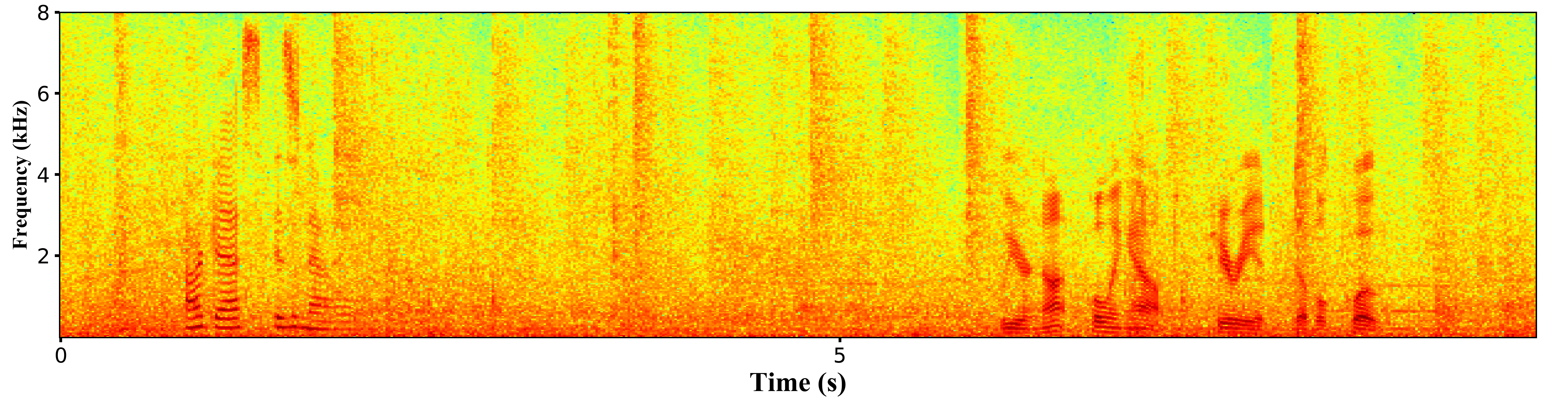}
		\end{minipage}
	}
	\subfloat[]{
		\begin{minipage}{9cm}
			\centering
			\includegraphics[width=1\textwidth]{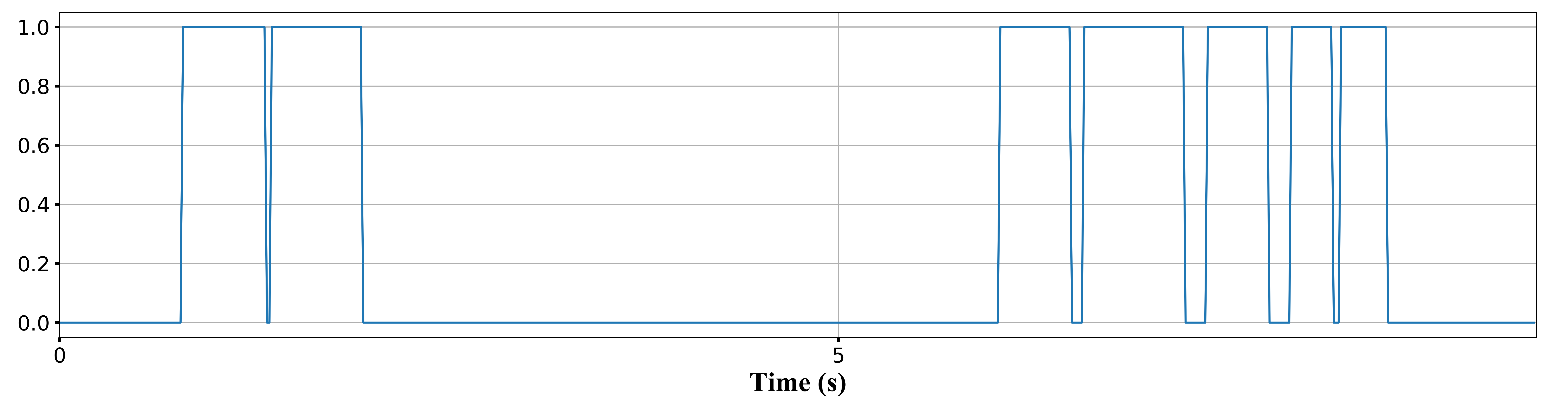}
		\end{minipage}
	}
	\\
	\subfloat[]{
		\begin{minipage}{9cm}
			\centering
			\includegraphics[width=1\textwidth]{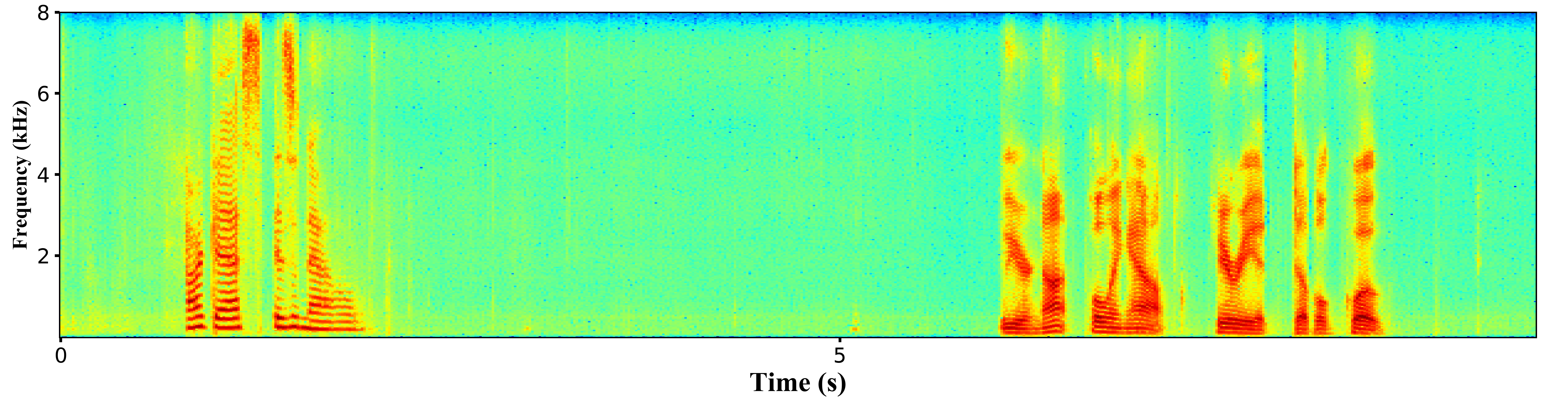}
		\end{minipage}
	}		 
	\subfloat[]{
		\begin{minipage}{9cm}
			\centering
			\includegraphics[width=1\textwidth]{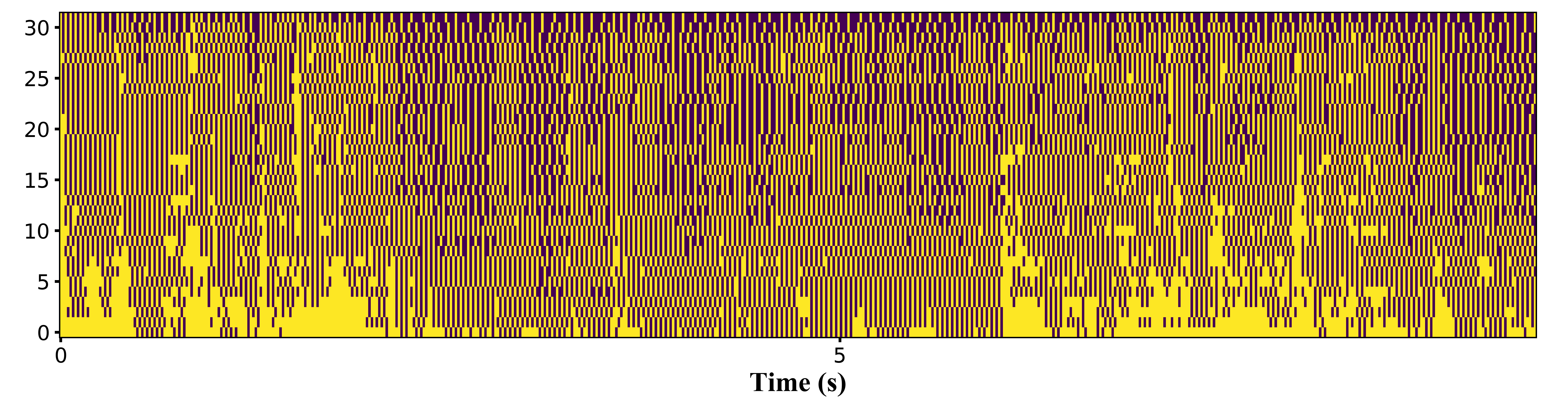}
		\end{minipage}
	}
	\\
	\subfloat[]{
		\begin{minipage}{9cm}
			\centering
			\includegraphics[width=1\textwidth]{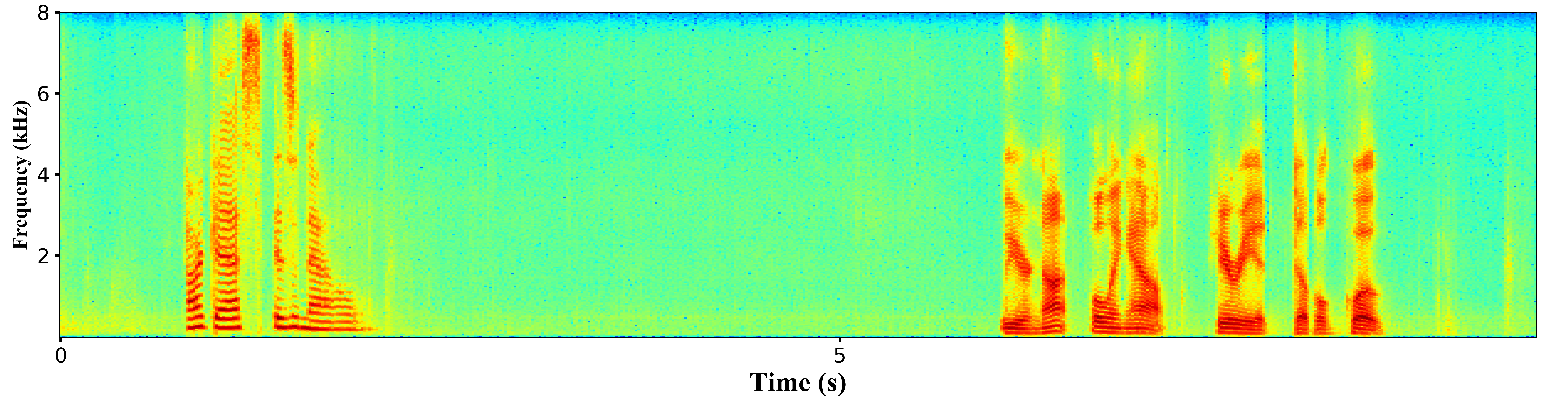}
		\end{minipage}
	}	
	\subfloat[]{
		\begin{minipage}{9cm}
			\centering
			\includegraphics[width=1\textwidth]{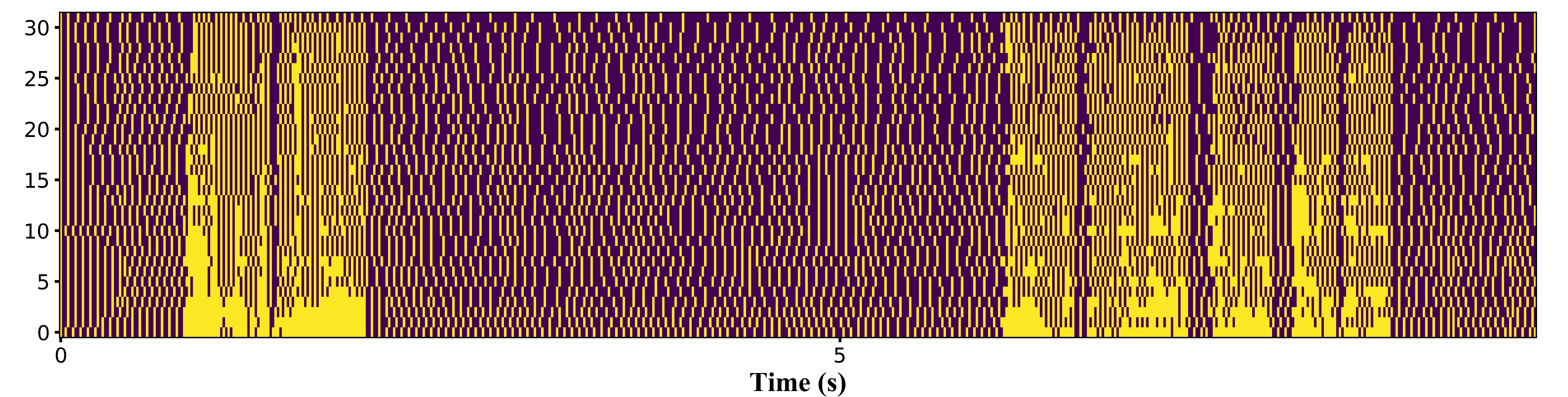}
		\end{minipage}
	}	 
	\caption{Spectrogram visualization and binary gate visualization of the last Skip-inter-RNN.
		(a) Noisy speech. (b) The VAD labels of the clean speech. (c) Enhanced speech by DPCRN-All-skip. (d) The binary gate visualization of the last Skip-inter-RNN. (e) Enhanced speech by DPCRN-All-skip with the VAD-guided skpping strategy. (f) The binary gate visualization of the last Skip-inter-RNN with the VAD-guided skpping strategy.}
	
	\label{fig:figure7}
\end{figure*}

Fig.~\ref{fig:figure7} shows the spectrogram processed by DPCRN-All-Skip ($\alpha = 1 \times 10^{-4}$) on a noisy sample with two utterances and long pure noise segments. The binary gate \(\mathbf{g}_{3;2}\) of the last Skip-inter-RNN is also visualized as a 2-D map in Fig.~\ref{fig:figure7}(d), where the horizontal and the vertical axes respectively denote time and the index of parallel RNNs. The bright bins in the map represent the updated states of RNNs and the dark bins represent the skipped states. From the map and the noisy spectrogram, we can observe the similarity between the distribution of the bright bins and the spectral envelope of the noisy speech. At noise segments, the hidden states are skipped significantly more frequently than the voice segments, which illustrates why the model “slacks off” at more tractable scenarios. Inspired by this, we can reduce the update rate at pure noise frames to further save the computational load while hold the original update rate at voice frames. Guided by the ideal VAD label shown in Fig.~\ref{fig:figure7}(b), the probability rescaling factor \(\gamma\) is set to 0.4 at pure noise frames and 1 at voice frames. Then we use Eq.(\ref{eq20}) to control the update rate. The results shown in Fig.~\ref{fig:figure7}(e)(f) illustrate that the Skip-RNNs skip more frequently at pure noise segments without spectrogram artifacts.

\begin{figure*}[ht]
	\centering
	\subfloat[]{
		\begin{minipage}{8cm}
			\centering
			\includegraphics[width=1\textwidth]{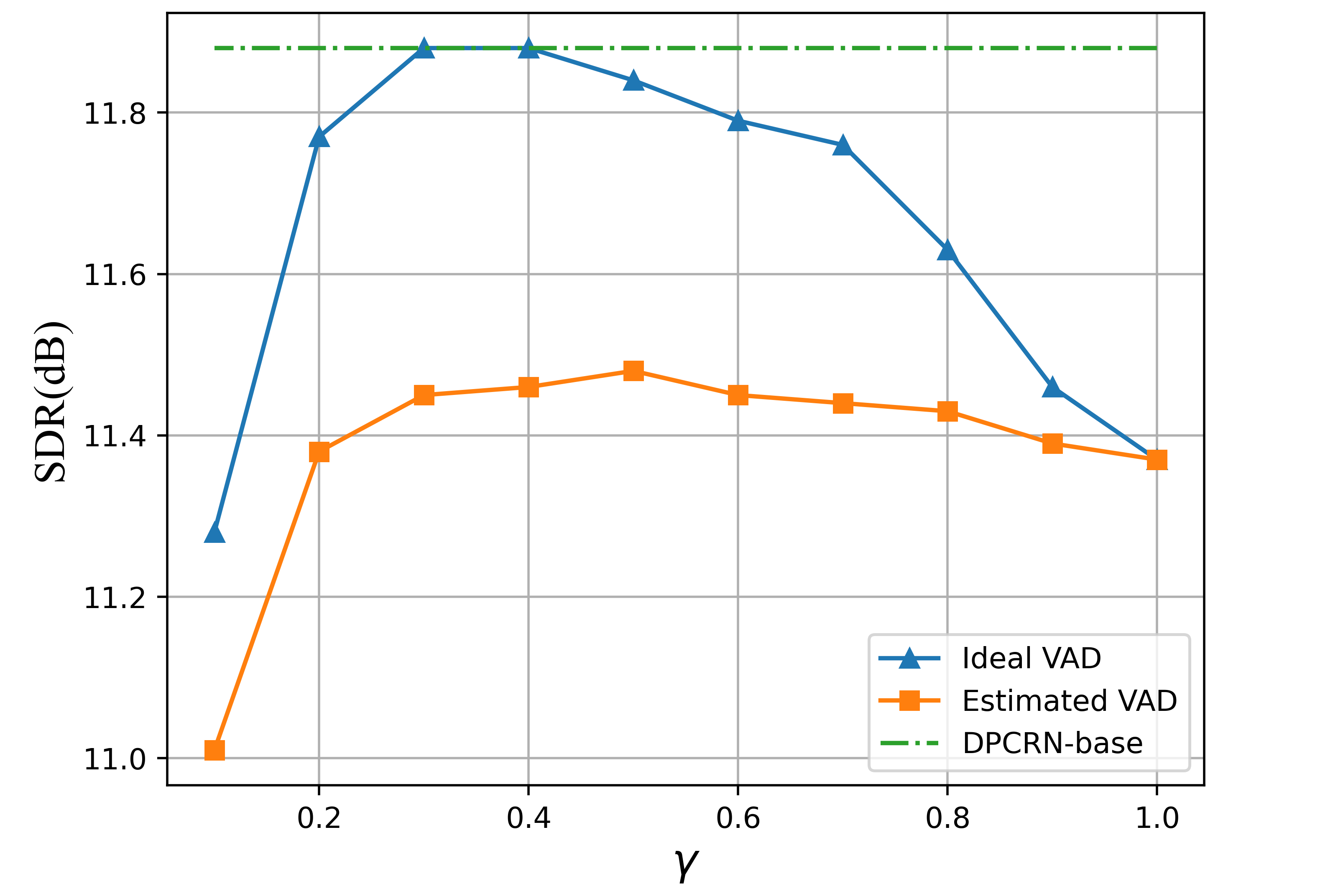}
		\end{minipage}
	}
	\subfloat[]{
		\begin{minipage}{8cm}
			\centering
			\includegraphics[width=1\textwidth]{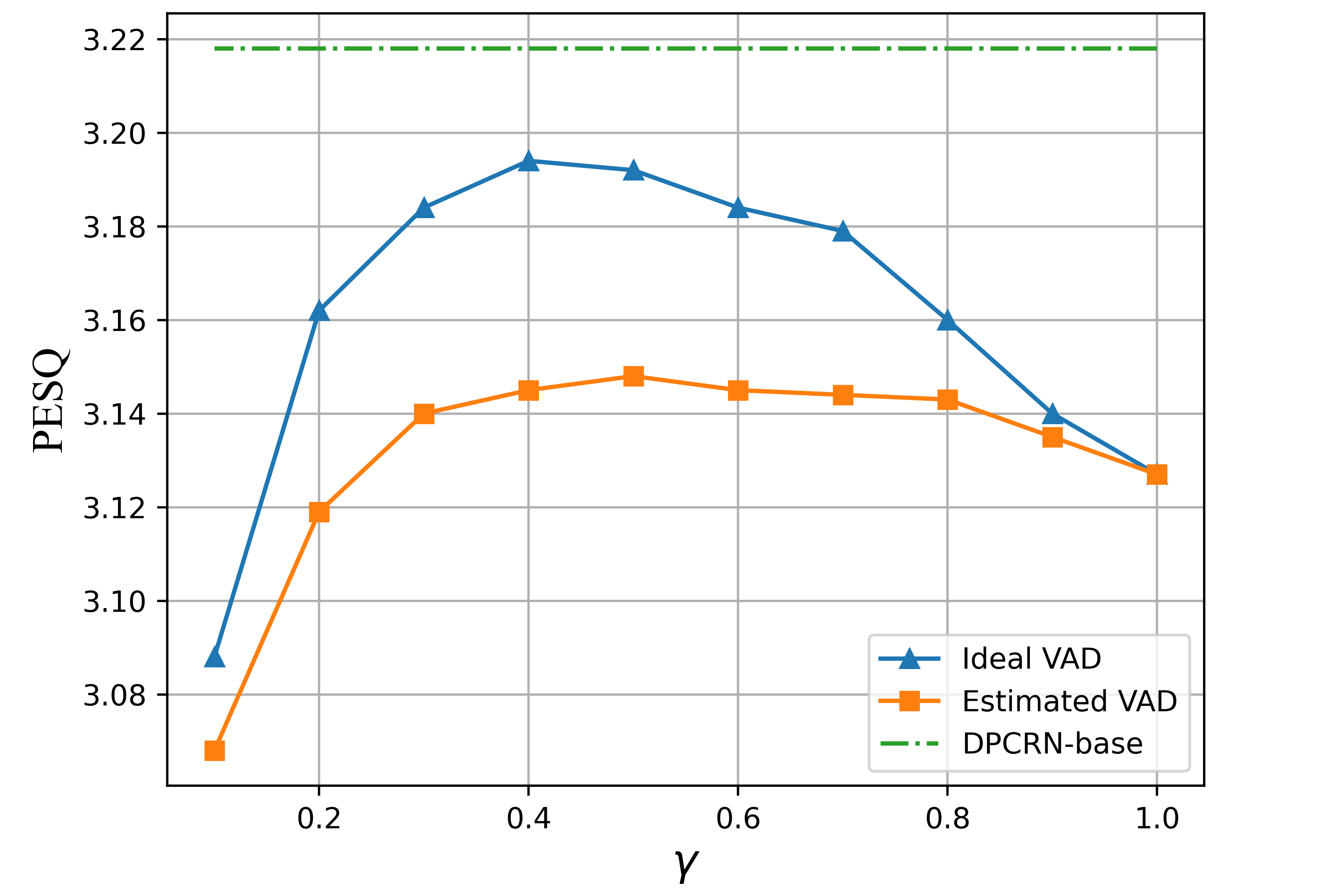}
		\end{minipage}
	}
\\  
	\subfloat[]{
		\begin{minipage}{8cm}
			\centering
			\includegraphics[width=1\textwidth]{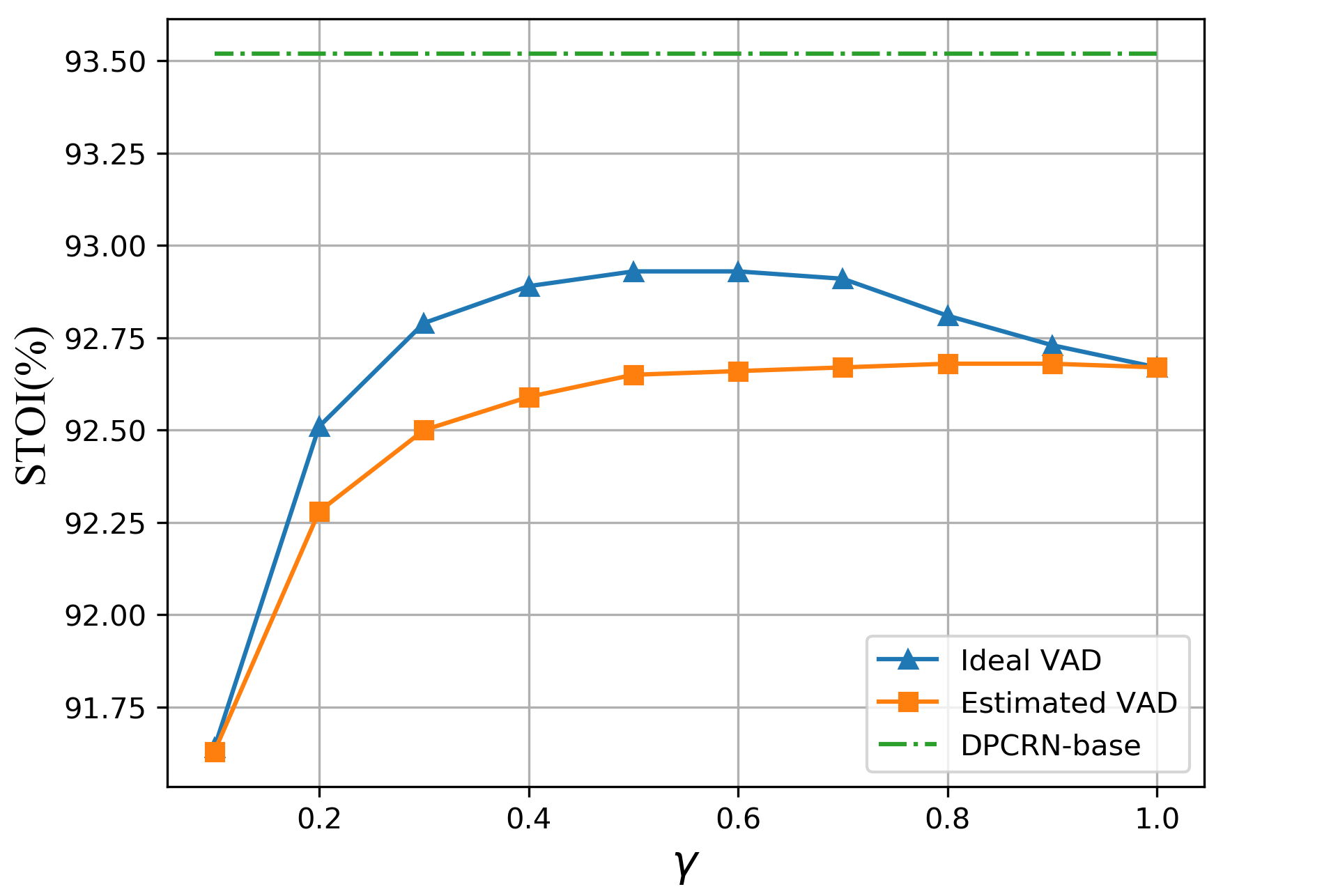}
		\end{minipage}
	}	
	\subfloat[]{
		\begin{minipage}{8cm}
			\centering
			\includegraphics[width=1\textwidth]{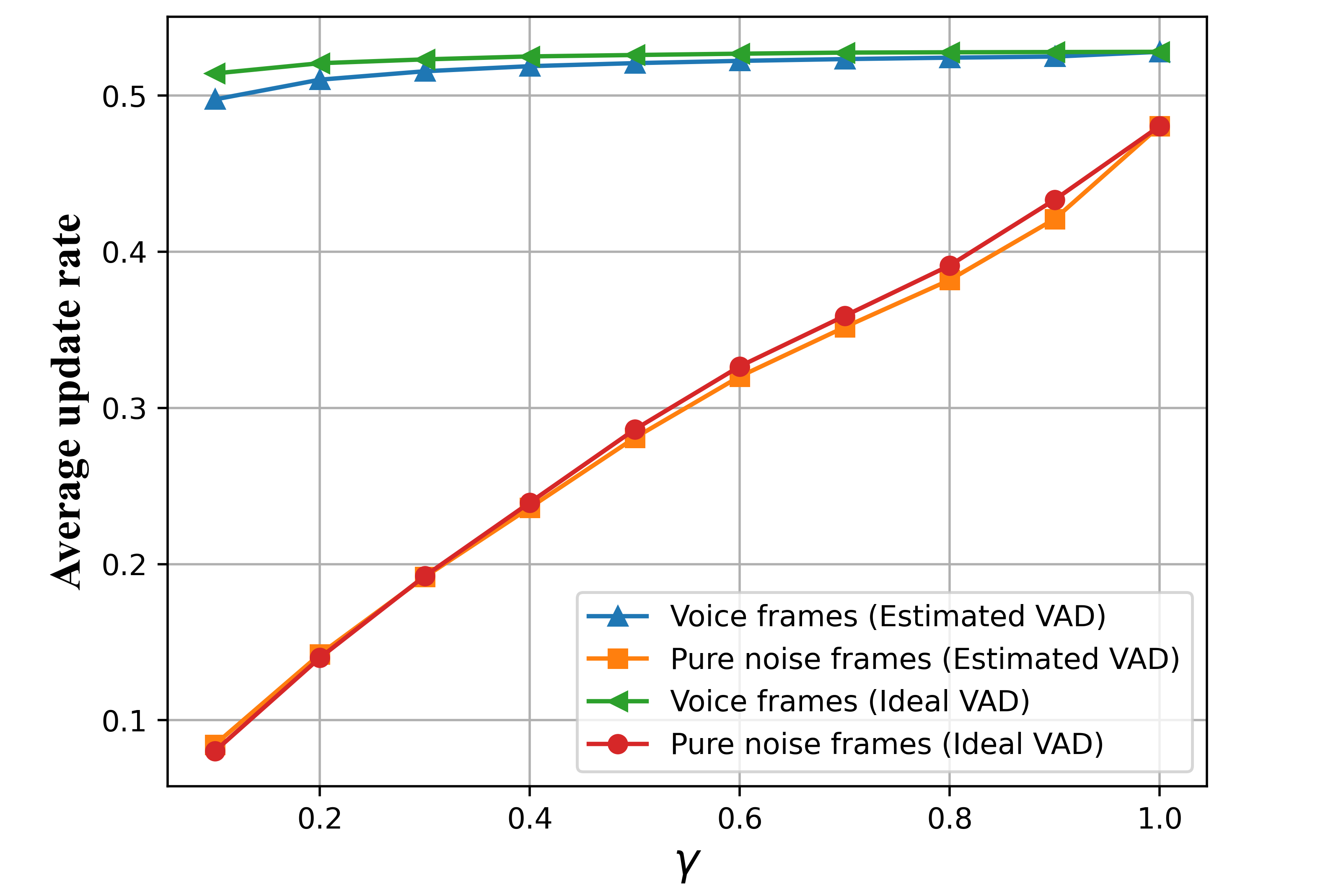}
		\end{minipage}
	}
	\caption{Results of the VAD-guided skipping strategy on the long audio test set. 
		(a) SDR scores, (b) PESQ scores and (c) STOI scores at different \(\gamma\). (d) The average update rates on voice and pure noise frames at different \(\gamma\).}
	
	\label{fig:figure8}
\end{figure*}

We test the VAD-guided skipping strategy on the long audio test set. DPCRN-All-skip is trained with a VAD estimator and the target average update rate is set to 0.5. We apply different \(\gamma\) only at pure noise frames and the objective results are shown in Fig.~\ref{fig:figure8}(a)(b)(c). As expected, the performance deteriorates significantly when \(\gamma\) is less than 0.3. However, all the metrics with the ideal VAD labels do not show a monotonic increasing trend, and the model even achieves the same SDR score as DPCRN-base (shown as the green dotted lines) when \(\gamma=0.4\), because pure noise frames can be more effectively attenuated\footnote{Exemplary audio samples can be found at https://github.com/Le-Xiaohuai-speech/SKIP-DPCRN-samples}. Using the estimated VAD labels leads to weaker performance than using the ideal labels, but the SDR and PESQ curves still show the benefit of small \(\gamma\). Overall, the model gets the best results when \(\gamma=0.5\) and the computational load of the DPRNN modules at pure noise segments is only 28\% of the baseline. Fig.~\ref{fig:figure8}(d) presents the average update rates with different \(\gamma\). The average update rates are almost constant at voice frames and demonstrate a close to linear relationship with \(\gamma\) at pure noise frames, which allows the model to preserve speech while further reduce the computational load. The accuracy of the VAD estimator on the long audio test set is 88.2\%. Note that the VAD estimator is not the focus of this paper and in practice better estimators may lead to higher performance.  

\linespread{1.15}
\begin{table*}[h]
	\caption{Performance of CRUSE and GCRN on two test sets sets.}
	\label{tab:table4}
	\centering
	\begin{tabular}{c c cccc c cccc}
		\toprule
		Test set    & \textbf{} & \multicolumn{4}{c}{NOISEX} & \textbf{} & \multicolumn{4}{c}{MUSAN} \\
		\cline{1-1}  \cline{3-6} \cline{8-11} 
		Metrics     &      & SDR (in dB) & PESQ          & STOI (in \%)   & Update rate      &     & SDR (in dB)   & PESQ           & STOI (in \%)   & Update rate \\
		\midrule
		
		Noisy       &      & 0.091       & 1.433         & 74.94          & -                &     & 0.043         & 2.004          & 80.76          & - \\
		
		CRUSE-base  &      & 8.423       & 2.457         & 85.65          & 1.00            &     & 10.29         & \textbf{2.834} & 91.41          & 1.00 \\	
		
		CRUSE-skip  &      & 8.426       & \textbf{2.457}& 85.58          & \textbf{0.496}  &     & 10.25         & 2.822          & 91.27          & \textbf{0.491}\\
		
		GCRN-base   &      & \textbf{9.393}  & 2.359     & \textbf{87.32} & 1.00            &     & \textbf{11.19}& 2.654          & \textbf{91.58} &1.00 \\
		
		GCRN-skip   &      & 9.348       & 2.364         & 87.07          & 0.697           &     & 11.14         & 2.650          & 91.36           & 0.642  \\
		\bottomrule
	\end{tabular}
\end{table*}

\subsection{Results on CRUSE and GCRN}

Table~\ref{tab:table4} presents the results of CRUSE and GCRN, where CRUSE-base and GCRN-base are the baseline models using the conventional RNN, and CRUSE-skip and GCRN-skip use Skip-RNN. It can be seen that after applying Skip-RNN, the two models remain the comparable performance to the baseline models with a significantly lower update rate, which indicates that Skip-RNN can be fused into any model that use parallel RNN structure. The final update rates of GCRN-skip is higher than CRUSE-skip because the former utilizes fewer parallel RNNs with a higher hidden dimension, which makes the model harder to optimize. In addition, the audio artifacts described in \cite{ref20} are not observed in all experiments, which is mainly due to the parallel RNNs and the CED structure. The parallel RNNs reduce the impact of the skipping on the updates of the output and the detailed information transferred from the encoder by the CED structure helps further recover the speech.

\section{Conclusion}
Parallel RNNs have been proven to be an important module for several high performance speech enhancement models. In this paper, we find that it is effective to introduce the Skip-RNN strategy into these models since the RNNs can be updated intermittently without interrupting the update of the output mask. Therefore the computational burden can be significantly reduced without evident audio artifacts. Using DPCRN as an example, we investigate several regularization strategies to induce more skipped states and introduces a VAD estimator to better exploit the characteristics of the noisy spectrogram. The experimental results demonstrate the superiority of the proposed model over those from the commonly used compression strategies, i.e., directly training a smaller model and structured pruning. Experiments on CRUSE and GCRN also validate the generalization of the proposed inference skipping strategy.

\bibliographystyle{IEEEtran}
\bibliography{mybib}

\end{document}